\documentclass[aps,article,nofootinbib,groupedaddress]{revtex4}
\usepackage{graphicx}
\usepackage{amsmath,amssymb}
\usepackage{hyperref}
\usepackage[usenames]{color}

\hypersetup{
    colorlinks=true,
    linkcolor=red,
    citecolor=blue,
}

\newcommand{\be}{\begin{equation}}
\newcommand{\ee}{\end{equation}}
\newcommand{\bq}{\begin{eqnarray}}
\newcommand{\eq}{\end{eqnarray}}
\newcommand{\bsq}{\begin{subequations}}
\newcommand{\esq}{\end{subequations}}
\newcommand{\bc}{\begin{center}}
\newcommand{\ec}{\end{center}}

\newcommand{\gsim}{\raise.3ex\hbox{$>$\kern-.75em\lower1ex\hbox{$\sim$}}}
\newcommand{\lsim}{\raise.3ex\hbox{$<$\kern-.75em\lower1ex\hbox{$\sim$}}}

\begin{document}

\title{Scaling configurations of cosmic superstring networks and their cosmological implications}

\author{A.~Pourtsidou\footnote{E-mail: alkistis@jb.man.ac.uk}$^{a,b}$, A.~Avgoustidis\footnote{E-mail: a.avgoustidis@damtp.cam.ac.uk}$^{c}$, E.~J.~Copeland\footnote{E-mail: ed.copeland@nottingham.ac.uk}$^{b}$, L.~Pogosian\footnote{E-mail: levon@sfu.ca}$^{d}$ and D.A.~Steer\footnote{E-mail: steer@apc.univ-paris7.fr}$^{e}$}

\affiliation{$^a$Jodrell Bank Centre for Astrophysics, School of Physics and Astronomy, University of Manchester, Manchester M13 9PL, UK\\
$^b$School of Physics and Astronomy, University of Nottingham, University Park, Nottingham NG7 2RD, UK\\
$^c$Centre for Theoretical Cosmology, DAMTP, CMS, Wilberforce Road, Cambridge CB3 0WA, UK\\ 
$^d$Department of Physics, Simon Fraser University, Burnaby, BC, V5A 1S6, Canada \\
$^e$APC\footnote{Universit\'e 
Paris-Diderot, CNRS/IN2P3,  
CEA/IRFU and Observatoire de Paris}, 10 rue Alice Domon et L\'eonie Duquet,
 75205 Paris Cedex 13, France}

\date{\today}

\begin{abstract}
We study the cosmic microwave background temperature and polarisation spectra sourced by multi-tension cosmic superstring networks. First we obtain solutions for the characteristic length scales and velocities associated with the evolution of a network of F-D strings, allowing for the formation of junctions between strings of different tensions. We find two distinct regimes describing the resulting scaling distributions for the relative densities of the different types of strings, depending on the magnitude of the fundamental string coupling $g_s$. In one of them, corresponding to the value of the coupling being of order unity, the network's stress-energy power spectrum is dominated by populous light F and D strings, while the other regime, at smaller values of $g_s$, has the spectrum dominated by rare heavy D strings. These regimes are seen in the CMB anisotropies associated with the network.  We focus on the dependence of the shape of the B-mode polarisation spectrum on $g_s$ and show that measuring the peak position of the B-mode spectrum can point to a particular value of the string coupling.
Finally, we assess how this result, along with pulsar bounds on the production of gravitational waves from strings, can be used to constrain a combination of $g_s$ and the fundamental string tension $\mu_F$.  Since CMB and pulsar bounds constrain different combinations of the string tensions and densities, they result in distinct shapes of bounding contours 
in the $(\mu_F, g_s)$ parameter plane, thus providing complementary constraints on the properties of cosmic superstrings. 
\end{abstract}

\maketitle

\section{Introduction}
\label{intro}

Although it has been established that a network of cosmic strings cannot source the majority of the observed cosmic microwave background (CMB) temperature anisotropy \cite{Albrecht:1997nt}, the CMB can still provide a distinctive signature of their presence through the specific primordial B-mode polarisation spectrum \cite{Seljak:1997ii,Battye:1998js,PogosianTye,SelSlo,Pogosian:2007gi,Bevis:2007qz,Urrestilla:2008jv,Mukherjee:2010ve}. The spectrum generated by strings is different from the one generically produced from tensor modes arising in inflationary scenarios, and future probes of the B-mode should be able to reveal the presence of cosmic strings,  even if strings contribute as little as $0.1\%$ to the CMB temperature anisotropy \cite{SelSlo,Pogosian:2007gi,Bevis:2007qz,Urrestilla:2008jv,Mukherjee:2010ve}.

Interest in cosmic strings has revived following the realisation that they can arise in superstring theory \cite{Sarangi:2002yt, Jones:2003da}, for example in models of brane inflation \cite{Dvali:1998pa,models,Kachru:2003sx,Burgess:2004kv} (for a review see \cite{Copeland:2009ga}). Cosmic superstrings can have small tensions ($10^{-12} \gsim G\mu \gsim 10^{-7}$ \cite{Sarangi:2002yt, Jones:2003da, Polchinski:2004ia}), can be effectively stable over cosmological timescales, and can stretch over cosmological distances \cite{Copeland:2003bj,Leblond,Dvali:2003zj}. Hence, they can have interesting cosmological implications. Furthermore, their intercommutation probabilities can be significantly less than unity \cite{Jones:2003da, Jackson:2004zg, Hanany:2005bc,Jackson:2007hn} and, because of the charges present on them, they can zip together to form Y-junctions (trilinear vertices), leading to more complicated networks than those usually considered in the case of `standard' Abelian cosmic strings. Understanding the imprint of such additional network features on observables, such as CMB temperature and polarisation, is a step that may lead to interesting new constraints on the basic parameters of string theory, such as the string coupling $g_s$ and the fundamental string tension $\mu_F$.

Several approaches have been developed to model the evolution of cosmic string networks, and an interesting recent attempt to extend them to cosmic superstring networks -- which contain different types of string -- is due to Tye, Wasserman and Wyman \cite{TWW}. Their model, based on the ``velocity dependent one-scale'' model of Martins and Shellard \cite{VOS,VOSk}, describes evolution of a multiple tension string network (MTSN) under the assumption that all types of strings have the same correlation length and root-mean-square velocity. By studying the evolution of the number density of strings, they find that scaling is achieved when the energy associated to the formation of junctions is assumed to be radiated away. This model has been extended in \cite{NAVOS}, where the authors assigned a different correlation length and velocity to each string type, and enforced energy conservation at each junction. Scaling is again achieved (with different number densities), but not as generically as in \cite{TWW}. 

In a complementary approach, a number of authors have studied the kinematics of cosmic string collisions \cite{CopKibSteer1,CopKibSteer2}. When two Nambu-Goto (NG) strings (of generally different tensions) collide, rather than intercommuting in the standard way, they can form two junctions and a linking string of a third tension. Kinematically this can only occur if the relative orientation, velocity and string tensions lie in certain ranges. In \cite{CopFirKibSteer}, the authors extended their earlier studies to $(p,q)$-cosmic superstrings by modifying the NG equations to take into account the additional requirements of flux conservation. Once again the kinematic conditions required for the formation of Y-junctions were established, with results very similar to the ones obtained for NG strings.  These kinematic constraints have been checked quite extensively with dynamical field theory simulations of strings collisions, and the agreement is (generally) good \cite{Sakellariadou:2008ay,Salmi,Urrestilla:2007yw,Bevis:2008hg,Bevis:2009az}.  Recently, in \cite{Avgoustidis:2009ke} these constraints have been incorporated into the generalised MTSN velocity one-scale model of \cite{NAVOS}, leading to new conditions required for scaling and thereby providing the most complete model of cosmic superstring evolution to date. 

In this paper, we use the model of \cite{Avgoustidis:2009ke} to study the evolution of a cosmic superstring network for different values of the string coupling $g_s$ and different charges $(p,q)$ on the strings. We find that in all cases the three lightest strings, i.~e. the $(1,0)$, $(0,1)$ and $(1,1)$ strings, dominate the string {\it number} density. When the string coupling is large, $g_s \sim {\cal O}(1)$, most of the network {\it energy} density is in the lightest $(1,0)$ and $(0,1)$ strings (respectively F and D strings), whose tensions are approximately equal. At smaller values of $g_s \sim {\cal O}(10^{-2})$, the $(1,0)$ string becomes much lighter than both the  $(0,1)$ and $(1,1)$ strings, and dominates the string {\it number} density. However, because of their much larger tension, the {\it energy} density of the network at small couplings can be dominated by the rarer $(0,1)$ and $(1,1)$ strings. 
The existence of these two distinct limiting scaling behaviours at large or small values of $g_s$ is quite generic, although the specific details are somewhat dependent on the model-dependent value of the effective volume of the compactified dimensions. In either of the two limiting regimes, the {\it energy} density of the multi-tension network is effectively dominated by strings of one tension.

With the scaling solutions to hand we then focus on the CMB imprints of these networks, using a modified version of the publicly available code CMBACT \cite{Pogosian:1999np, cmbact}. In particular, we extend the Unconnected Segment Model (USM), first introduced in  \cite{Albrecht:1997nt,ABR99}, to describe the MTSN of \cite{Avgoustidis:2009ke} and implement it in CMBACT to obtain the CMB temperature and polarisation spectra. We find that for sufficiently large values of the parameter $w$, which is inversely proportional to the effective volume of the compactified dimensions (see Eq.~(\ref{wpar}) below), the two limiting regimes, one with the network energy dominated by light populous strings and the second with it dominated by rare heavy strings, can each produce distinct shapes for their CMB B-mode polarisation. In particular, for $w \sim 1$, the position of the peak in the B-mode spectrum is at $\ell \approx 770 $ for $g_s=0.9$ and at $\ell \approx 610$ for $g_s=0.04$. This opens up the exciting possibility that upcoming observations may not only constrain the overall contribution of strings, but in fact rule out certain values of the string coupling.  Namely, the combination of the normalisation and the peak position of the B-mode spectrum can point to a particular combination of $g_s$ and the fundamental string tension $\mu_F$.

It is common to report constraints on standard cosmic strings in terms of bounds on the single dimensionless string tension $G\mu$. These bounds have an implicit assumption on the number density of strings corresponding to the usual Abelian Higgs model strings with intercommutation probability ${\cal P} =1$.   However, in a more general situation such as that of cosmic superstrings considered here, there can be smaller intercommutation probabilities as well as strings of different tensions. As a result, each type of string will in principle have a different number density: the same fraction of CMB anisotropy can be sourced either with many light strings or with a few heavy ones. In general, each type of observational bound will constrain a different combination of the string tensions and densities (which, for cosmic superstrings, are derived from the fundamental string tension $\mu_F$ as well as $g_s$). In particular CMB and pulsar bounds, which we discuss in Section~\ref{sec:pulsars}, will lead to different shapes of bounding contours in the $(\mu_F, g_s)$ parameter plane. We show that combining these two constraints can lead to complementary constraints on properties of superstrings. The position of the peak in the B-mode spectrum can be used to further eliminate a large region of the $(\mu_F, g_s)$ parameter space.

This paper is organised as follows. In Section \ref{sec:scaling} we summarise the extended VOS model which describes multi-tension networks with junctions, and for the case of cosmic superstrings we present the scaling solutions as a function of $g_s$. In Section \ref{sec:cmb} we determine the temperature and B-mode spectra for these scaling solutions using a generalised version of CMBACT. In Section \ref{sec:pulsars} pulsar constraints on gravitational waves from string networks are discussed, and we conclude in Section \ref{sec:conc}.

\section{Scaling of cosmic superstring networks}
\label{sec:scaling}

Cosmic superstrings provide an example of a network of strings with multiple tensions, and one in which the strings can join at Y-shaped junctions. A general $(p,q)$ cosmic-string has $p$ quanta of F charge (associated with fundamental strings) and $q$ quanta of D charge (carried by D-branes) \cite{Schwarz:1995dk,Witten:1995im,Polchinski:2004ia}.

We begin this section by briefly reviewing the velocity dependent one-scale (VOS) model for single-tension strings, before generalising it (section \ref{multimu}) to a model for a network of $N$ different types of strings with junctions following \cite{NAVOS,Avgoustidis:2009ke}. We then customise the parameters of this model to the case of $(p,q)$ superstrings (section \ref{superparams}) paying particular attention to their dependence on the string coupling $g_s$. While the exact values of the parameters are model-dependent, e.g. depend on the choice of the compactification manifold, we are able to identify general trends in their dependence on $g_s$. These trends, in turn, lead to two distinct scaling scenarios in the limits of large and small $g_s$ (section \ref{scaling_gs}) that may be distinguished observationally. To gain additional intuition, we have included an Appendix where we analytically solve for the scaling parameters in the limit of small $g_s$ to complement the exact numerical solutions.

\subsection{The VOS model for a network of a single type of string with no junctions}
\label{sec:vos}

The so-called ``one-scale model" \cite{Kibble} assumes that the string network can be characterised by a single length scale, namely the correlation length $L$ defined by 
\be
\rho=\frac{\mu}{L^2} \ ,
\label{eq:rho}
\ee
where $\rho$ is the energy density in the infinite string network and $\mu$ is the string tension. In principle, there are at least two different fundamental length scales in the network --- the typical ``smoothness" length $L$ of long strings, and the average distance between strings $\bar{L}$. The one-scale model takes the two lengths to be equal ($L$=$\bar{L}$), an approximation which appears to be reasonably well satisfied \cite{Austin:1993rg} for a network of NG strings of tension $\mu$ with intercommutation probability ${\cal P}=1$.

As the network evolves, string-string and self-intersections result in formation of loops which subsequently decay. The energy loss rate can be approximated by \cite{Kibble}
\be
\dot{\rho}=-2\frac{\dot{a}}{a}\rho - \frac{\rho}{L},
\label{enlossrate}
\ee
where $\dot{} = d/dt$ and $a(t)$ is the scale factor. The first term accounts for the expansion of the universe,
and the second for string interactions with associated loop-formation.
The network evolves towards a scaling regime, in which $L$ is constant relative
to the horizon $d_H \sim t$. Indeed, setting $L(t)=\xi(t)t$, it follows from Eq.~(\ref{enlossrate}) that
\be
\frac{\dot\xi}{\xi}=\frac{1}{2t}\left(2(\beta-1)+\frac{1}{\xi}\right),
\label{gammaeq}
\ee
where $a(t) \sim t^{\beta}$ ($\beta={1}/{2}$ in a radiation era, $\beta={2}/{3}$ in a matter era).
The attractor scaling solution of Eq.~(\ref{gammaeq}) is
\be
\xi=[2(1-\beta)]^{-1}.
\label{scalingsol}
\ee

The ``velocity-dependent one-scale model'' (VOS) is slightly more sophisticated 
in that it introduces a dynamical velocity component $v$ to the equations, describing the root mean square (rms) velocity of string segments.  This model agrees quantitatively 
 with results obtained from NG numerical simulations  \cite{VOS,VOSk}. The 
 relevant equations are
\be
\dot{\rho}=-2\frac{\dot{a}}{a}(1+v^2)\rho - \frac{\tilde{c} \, v\rho}{L},
\label{vosrho}
\ee
\be
\dot{v}=(1-v^2)\left(\frac{k}{L}-2\frac{\dot{a}}{a}v \right),
\label{vosvel}
\ee
where the constant $\tilde{c}$ represents the efficiency of loop formation, and $k$ is the curvature parameter which indirectly encodes information about the small-scale structure on strings. It can be expressed as a function of the velocity \cite{VOSk}; 
\be
  k=\frac{2\sqrt{2}}{\pi}\left(\frac{1-8v^6}{1+8v^6}\right),
  \label{k-eqn}
\ee 
which incorporates the Virial condition $v^2 \leq \frac{1}{2}$, observed in simulations.
Following the same procedure as before, one now finds a scaling solution with
\bq
\xi^2&=&\frac{k(k+\tilde{c})}{4\beta(1-\beta)} \,,
\label{vosscaling1}
\\
v^2&=&\frac{k(1-\beta)}{\beta(k+\tilde{c})} \,.
\label{vosscaling2}
\eq
Of particular note for the following is that the velocity 
also enters a scaling regime (\ref{vosscaling2}) in which it stays constant in time. 

The relation between intercommutation probability ${\cal P}$ and the loop chopping efficiency parameter $\tilde c$ is not fully understood at present. Nambu-Goto simulations of strings interacting with a microphysical probability ${\cal P}<1$ suggest that $\tilde c \simeq {\cal P}^{1/3}$ in both the matter and radiation era \cite{Avgoustidis:2005nv}. A different dependence, $\tilde c \simeq {\cal P}^{1/2}$, was reported in \cite{Sakellariadou:2004wq} based on a flat space simulation. In the subsequent sections we follow \cite{Avgoustidis:2005nv} and take $\tilde c$ to scale as the cubic root of the corresponding intercommutation probability. Such a weak dependence of $\tilde c$ on ${\cal P}$ can be attributed to the presence of small scale structure on long strings, allowing for multiple chances of intercommutation when two string segments cross \cite{Avgoustidis:2005nv}.

\subsection{Evolution of multi-tension networks with junctions}\label{multimu}

To describe the evolution of a multi-tention string network (MTSN) with junctions we adopt the model developed in \cite{NAVOS,Avgoustidis:2009ke} in which we solve for the energy densities and rms velocities of each string type using the following set of equations:
\be\label{rho_idtgen} 
    \dot\rho_i = -2\frac{\dot a}{a}(1+v_i^2)\rho_i-\frac{c_i 
    v_i\rho_i}{L_i} - \sum_{a,k} \frac{d_{ia}^k   
    \bar v_{ia} \mu_i \ell_{ia}^k(t)}{L_a^2 L_i^2} + \sum_{b,\,a\le b}   
    \frac{d_{ab}^i \bar v_{ab} \mu_i   
    \ell_{ab}^i(t)}{L_a^2 L_b^2}         \, ,
  \ee  
  \bq
    \dot v_i = (1-v_i^2)\left[\frac{k_i}{L_i}-2\frac{\dot a}{a}v_i 
+\sum_{b,\,a\le b} b_{ab}^i \frac{\bar  
    v_{ab}}{v_i}\frac{(\mu_a+\mu_b-\mu_i)}{\mu_i}\frac{\ell_{ab}^i(t)  
    L_i^2}{L_a^2 L_b^2}\right] \, . \label{v_idtgen}     
\eq
Here $\mu_i$ is the tension of the $i$th type of string, and, as in the case of single tension networks, we define a correlation length $L_i$ through
\be
\rho_i =\frac{\mu_i}{L_i^2} \ .
\ee

In analogy to the single string case, the coefficients $c_i$ in (\ref{rho_idtgen}) quantify the efficiency with which self-interactions of strings of type $i$ produce closed loops, thus removing energy from the long string network. The last two terms in Eq.~(\ref{rho_idtgen}) model the effect of collisions between strings of different types, leading to  the formation of new segments ending on 3-string junctions. In particular, the penultimate term describes the loss of energy, from network $i$, due to string segments of type $i$ colliding with segments of type $a$ and forming links of type $k$.  Similarly, the last term models the energy gain in network $i$ through collisions between different strings $a$ and $b$, leading to the formation of a link of type $i$.  The parameter $d_{ij}^k=d_{ji}^k$, which we will discuss in more detail below, is essentially the probability with which strings of types $i$ and $j$ interact and produce a type $k$ segment. This parameter captures interactions at the quantum level and incorporates volume effects \cite{Jones:2003da,Jackson:2004zg}, as well as the kinematic constraints discussed in \cite{CopKibSteer1,CopFirKibSteer,Salmi,Avgoustidis:2009ke}.  The average length of the links formed by this process at time $t$ is denoted by $\ell_{ij}^k(t)$, whose explicit form will be given below (Eq.~(\ref{zipper-length})). In Eq.~(\ref{v_idtgen}), the coefficients $k_i$ are curvature parameters which indirectly encode information about the small-scale structure on the strings.  We will follow \cite{VOSk} and take them to depend on the rms velocities as
\be
  k_i=\frac{2\sqrt{2}}{\pi}\left(\frac{1-8v_i^6}{1+8v_i^6}\right) \, .
  \label{ki-eqn}
\ee 
The parameters $b_{ab}^i$ have been introduced in order to interpolate between the model of \cite{NAVOS} (where $b_{ab}^i = d_{ab}^i$), in which the energy liberated by the formation of junctions is redistributed in the network as kinetic energy, and a model analogous to that in \cite{TWW} (corresponding to $b_{ab}^i =0$), in which all of this energy is assumed to be radiated away. It is more likely that realistic situations are somewhere inbetween, with $b_{ab}^i < d_{ab}^i$, corresponding to a fraction of the liberated energy being redistributed, the rest being radiated away. Finally, $\bar v_{ab}$ is the magnitude of the relative velocity between strings of type $a$ and $b$ averaged over all directions, that is $\bar v_{ab}=\sqrt{v_a^2+v_b^2}$.

\subsection{Parameters for cosmic superstring networks}
\label{superparams}

In cosmic superstring networks, the tension of each string type is determined by the corresponding charges, $(p,q)$, and the string coupling $g_s$. In flat spacetime \cite{Schwarz:1995dk,Witten:1995im,Polchinski:2004ia},
\be\label{pqtension} 
 \mu_i \equiv  \mu_{(p_i,q_i)}=\frac{\mu_F}{g_s} \sqrt{p_i^2 g_s^2+q_i^2} \ ,
\ee 
where $\mu_F$  is the tension of the lightest fundamental string (F string) carrying charge $(1,0)$. The D string has a charge $(0,1)$,  while the strings carrying charges $(p,q)$ with $p,q \ge 1$ can be thought of as bound states between $p$ F strings and $q$ D strings. There is an infinite hierarchy of such $(p,q)$ bound states, but, as was found in \cite{TWW,NAVOS,Avgoustidis:2009ke}, the cosmological evolution of interacting networks of this type lead to solutions in which only the first few lightest 
strings dominate, allowing us to truncate the system at finite $N$.  For our simulations we take $N=7$ implying that we have seven different types of string carrying charges $(p_i,q_i)\equiv (p,q)_i$ with
\be\label{strings}
\{(p,q)_i\} = \{(1,0),(0,1),(1,1),(2,1),(1,2),(3,1),(1,3)\} \ , \ \ (i=1,...,7) \,.   
\ee

The key parameters in Eqs.~(\ref{rho_idtgen}-\ref{v_idtgen}) are the self- and cross- interaction 
coefficients, $c_i$ and $d_{ij}^k$ respectively.  They are averaged network quantities that depend on the microphysical intercommuting probabilities of the corresponding interaction processes, which in turn can be modelled using string theory techniques \cite{Jackson:2004zg,Jackson:2007hn}. The cross-interaction coefficients $d_{ij}^k$ also depend on kinematic constraints on 3-string junctions \cite{CopKibSteer1,CopFirKibSteer,Salmi}.      

In the remainder of this subsection we discuss in detail the dependence of the $c_i$ and $d_{ij}^k$, as well as other parameters entering Eqs.~(\ref{rho_idtgen}) and (\ref{v_idtgen}), on the fundamental parameters of string theory such as $g_s$. Solutions of Eqs.~(\ref{rho_idtgen}) and (\ref{v_idtgen}) together with a discussion of the resulting network evolution is given in section \ref{scaling_gs}: a reader less interested in the following more technical discussion may turn directly to section \ref{scaling_gs}.

\subsubsection{Intercommuting probability ${\cal P}_{ij}$}

We begin by discussing the microphysical intercommuting probability ${\cal P}_{ij}$ arising from interactions between strings of type $i$ and $j$.  From ${\cal P}_{ij}$ one can determine both
the self- and cross- interaction coefficients, $c_i$ and $d_{ij}^k$, as discussed below.

For processes involving at least one F string, that is F-$(p,q)$ interactions, this probability ${\cal P}_{ij}$ can be calculated perturbatively in string theory \cite{Jackson:2004zg,Jackson:2007hn}.  The result is a function of the string coupling $g_s$, the relative velocity $v$ and angle $\theta$ of the incoming strings. However, there is also a model-dependent volume factor which depends on the size of the compact extra dimensions, parameterised by $w$ (which will be defined precisely later), and by the amplitude of fluctuations of the string position fields. The latter depends on the string tension and, therefore, on the string coupling $g_s$.  A convenient way to write the answer is
\be\label{PisFV}
{\cal P}_{ij}(v,\theta,w,g_s)={\cal F}_{ij}(v,\theta,g_s)\,{\cal V}_{ij}(w, g_s) \,,
\ee
where ${\cal F}_{ij}(v,\theta,g_s)$ is the volume-independent quantum interaction piece and 
${\cal V}_{ij}(w,g_s)$ encodes the volume dependence \cite{Jackson:2004zg,Jackson:2007hn}.  

\vspace{0.3cm}
\begin{center}
{\it 1a). $\; \;$ The volume-independent piece, ${\cal F}_{ij}(v,\theta,g_s)$}
\end{center}

As mentioned above, for interactions involving at least one F string, ${\cal F}_{ij}(v,\theta,g_s)$ corresponds to a perturbative amplitude that can be readily computed in string theory \cite{Jackson:2004zg}.  On the other hand, for interactions involving only D strings or heavier composites, the process is non-perturbative and less well understood: at present there are at least two approximate results, by Jackson et al.  \cite{Jackson:2004zg} (hereafter JJP)
based on a worldsheet calculation, and by Hanany \& Hashimoto \cite{Hanany:2005bc} (hereafter HH) using a field theory approach.  The two calculations are in good qualitative agreement, but there are quantitative differences reflecting the uncertainties arising from the currently incomplete understanding of such non-perturbative processes.  Nevertheless, these calculations provide a basis for systematically computing the network coefficients in Eq.~(\ref{rho_idtgen}-\ref{v_idtgen}), and allow us to study the effect of these uncertainties on the macroscopic evolution of the networks (section \ref{scaling_gs}).  As the available methods for calculating these processes improve (for recent progress see \cite{Jackson:2007hn}) we anticipate that such uncertainties will be eventually controlled.  Here, we shall use the results of JJP for the perturbative processes involving at least one F string, and the result of HH for the non-perturbative D-D interactions.  These have non-trivial dependencies on $v$ and $\theta$ and the relevant string charges $p$ and $q$ (for details see \cite{Jackson:2004zg,Hanany:2005bc}), but there are some key features with respect to their dependence on the string coupling $g_s$ which can be summarised: 
\begin{itemize}
\item F-F string interactions scale with $g_s^2$, 
\item F-$(p,q)$ interactions with $q\ge 1$ scale with $g_s$,
\item the non-perturbative D-D interactions scale with $U^{1/g_s}$, where $U$ is a number of order unity.
\end{itemize}
For interactions between heavier composites, i.e. $(p,q)$-$(p',q')$ with $q,q'> 1$ and $p,p'\ge 1$, the amplitude is not known but it is understood that it is enhanced with respect to the D-D amplitude by the multiplicity $qq'$ of the relevant Chan-Paton states \cite{Jackson:2004zg}.  In addition, for small values of the coupling $g_s$ we can neglect the effect of the light perturbative F strings so we will approximate the enhanced amplitude as ${\cal  F}_{(p,q)(p',q')}=1-(1-{\cal F}_{\rm DD})^{qq'}$.  The detailed form of the factors ${\cal F}_{ij}(v,\theta,g_s)$ assumed in this paper (including their full $v$, $\theta$ dependence) is shown in
table~\ref{TableFVs} (left column).

\begin{table}[h]
    \begin{center}
    \begin{tabular}{|c||c|c|} \hline
     Interaction ($ij$) & ${\cal F}_{\rm ij}$ & ${\cal V}_{\rm ij}$\\ \hline\hline
   F-F & $ \ g_s^2 \frac{\left( 1 -\cos\theta\sqrt{1-v^2}\right)^2}{8\sin\theta
     \, v \sqrt{1-v^2}}$ \ & $w$ \\ \hline 
   F-D & $ \ g_s \frac{v^2 + \left(\cos\theta\sqrt{1-v^2}\right)^2}{8\sin\theta
     \, v \sqrt{1-v^2}}$ \ & $ \ \min(w g_s^{-3/2},8w,1) \ $ \\ \hline  
   F-$(p,q)$\ , \ \  $q\ge 1$ & $ \ g_s \frac{q^2 v^2 + \left(g_s p - \cos\theta
        \sqrt{(1-v^2)(g_s^2p^2+q^2)}\right)^2}{8\sin\theta\, v 
      \sqrt{(1-v^2)(g_s^2p^2+q^2)}}\ \ $ & $ \ \min\left[(p^2 + q^2
    g_s^{-2})^{3/4}w, 8w, 1\right] \ $ \\ \hline 
   D-D & $ \ \min\left\{\frac{\sqrt{g_s}}{2\pi^{3/4}\theta^{3/4}}e^{2\sqrt{2/3}(\theta/v)}
    \exp\left[-\frac{4\sqrt{\pi}\theta^{3/2}}{g_s}e^{-4\sqrt{2/3}(\theta/v)}
    \right],1 \right\} \ $ & $\min(w g_s^{-3},1)$ \\ \hline 
   $(p,q)$-$(p',q')$\ , \ $q,q'\ge 1$ & $ \ 1-(1-{\cal F}_{\rm
       DD})^{qq'} \ $ & $ \ \min\left\{\left[(p^2 + q^2 g_s^{-2})(p'^2
     + q'^2 g_s^{-2})\right]^{3/4} w, 1 \right\} \ $ \\ \hline 
    \end{tabular}
    \end{center}
    \caption{\label{TableFVs} The coefficients ${\cal F}_{ij}$ and
      ${\cal V}_{ij}$ for different string interactions.}
   \end{table}

\vspace{0.3cm}
\begin{center}
{\it 1b). $\; \;$ The volume factors ${\cal V}_{ij}(w, g_s)$}
\end{center}

These volume factors (see Eq.~(\ref{PisFV})) arise because the strings are moving in a higher-dimensional space. Thus they can miss each other as they cross, leading to an overall suppression on the interaction amplitude that scales with the inverse of the volume of the extra dimensions. However, it has been argued \cite{Jackson:2004zg} that this suppression effect may not be as important as originally anticipated \cite{Jones:2003da}, because the string position fields are worldsheet scalars -- not protected by any symmetry -- and should therefore be stabilised at a minimum of a potential well (see, however, \cite{Avgoustidis:2007ju,O'Callaghan:2010sy}) rather than exploring the compact orthogonal dimensions.  In this case, there is still a volume effect arising from the fact that strings are quantum objects whose positions fluctuate around the classical minimum, thus giving rise to an effective volume that each string explores, which is, however, a small fraction of the total.  The size of the fluctuations is determined by the mass of the string, so this effective volume depends on the type of string and on the coupling $g_s$ (for details see \cite{Jackson:2004zg}).  For F-F interactions, both strings have the same tension and fluctuate by the same amount, but for F-D interactions (and for small $g_s$) the fluctuation of the heavier D string can be neglected, leading to a volume which is a factor of $(\sqrt{2})^6=8$ times smaller\footnote{Note that 6 is the number of extra dimensions.} than the corresponding F-F volume.  Finally, for D-D interactions the volume is approximately a factor $g_s^3=(\sqrt{g_s})^6$ smaller than the F-F volume \cite{Jackson:2004zg}.  The overall volume suppression factor becomes unity when the effective string volume becomes equal to the minimum ``string-scale volume" $V_{\rm min}=(2\pi^2\alpha')^3$.  Thus, defining our model-dependent parameter $w$ as  
\be\label{wpar}
w\equiv V_{\rm min}/V_{\rm FF}\in (0,1] \ , 
\ee
we have for the relevant suppression factors ${\cal V}_{\rm FF}=w$, ${\cal V}_{\rm FD}=\min(8w,1)$ and ${\cal V}_{\rm DD}=\min(w g_s^{-3},1)$ as long as $g_s\ll 1$. Here, we generalise these volume factors for $(p,q)$-strings and for $g_s\lesssim 1$ in a phenomenological way so as to reproduce the above limits. In particular, we assign a factor of $(p^2 + q^2 g_s^{-2})^{3/4}$ to each $(p,q)$-string. We show the resulting ${\cal V}_{ij}(w,g_s)$ in table~\ref{TableFVs} (right column).  This is simply a choice we make in order to be able to systematically calculate these suppressions for different string types, but is one which successfully reproduces the results of \cite{Jackson:2004zg} in the appropriate limits. The dependence of our network scaling results on this choice will be discussed in section \ref{scaling_gs} below. We are thus left with a residual model-dependent variable $w$ that we treat as an external tunable parameter like $g_s$.  Note that the choice $w\simeq 1$ for this parameter (corresponding geometrically to a compactification very close to the string scale) makes ${\cal V}_{ij}\simeq 1$ for all strings, so the dependence of the intercommuting probability ${\cal P}_{ij}$ on the string type and $g_s$ is determined only by the quantum interaction ${\cal F}_{ij}(v,\theta,g_s)$ in this case.                 
                
For the macroscopic string networks we are interested in, string collisions happen perpetually as the network evolves, with a range of relative velocities and angles.  We can therefore average out the velocity and angle dependence of the microphysical probabilities ${\cal P}_{ij}(v,\theta,w,g_s)={\cal F}_{ij}(v,\theta,g_s)\,{\cal V}_{ij}(w, g_s)$, simply by integrating over a Gaussian velocity distribution -- peaked at the relative scaling velocities\footnote{Note that this  
introduces an implicit dependence of the coefficients of Eqs. (\ref{rho_idtgen}-\ref{v_idtgen}) on the actual scaling 
solutions for the string velocities $v_i$.  This can be treated iteratively, by choosing initially some guess values 
for the scaling velocities to estimate these coefficients, and then update them iteratively using the returned scaling velocities until convergence.} -- and a flat distribution on the collision angles.  This yields the probabilities ${\cal P}_{ij}(w,g_s)$ which now depend only on our two free parameters $w$ and $g_s$.            

The next step is to turn these probabilities into the network coefficients $c_i$ and $d^k_{ij}$ appearing in the macroscopic evolution Eq.~(\ref{rho_idtgen}-\ref{v_idtgen}).

\subsubsection{Self-interaction 
coefficients $c_i$}

For {\it self-interactions}, numerical simulations of Nambu-Goto string networks with reduced microphysical probabilities \cite{Avgoustidis:2005nv}, suggest that the effective (loop-chopping efficiency) coefficient $\tilde c$ in Eq.~(\ref{vosrho}) scales with the third root of the microphysical probability ${\cal P}$ (refer to the discussion in section \ref{sec:vos}).  Thus we will take here
\be\label{cis}
c_i = {\tilde c} \times {\cal P}_i^{1/3} \, , 
\ee
where we have denoted ${\cal P}_{ii}\equiv {\cal P}_{i}$ and ${\tilde c}$ is chosen so as to reproduce single network results for ${\cal P}_i=1$, $d_{ij}^k=0$. Following \cite{VOS,VOSk}, we use ${\tilde c}={\tilde c}_r=0.23$ in the radiation era, and ${\tilde c}={\tilde c}_m=0.18$ in the matter era. 

\subsubsection{Cross-interaction coefficients $d_{ij}^k$}

 For the {\it cross-interactions} (i.e. with $i\ne j$) producing zipped configurations, there are at present no network simulations to compare to in order to determine the dependence of $d_{ij}^k$ on the microscopic probabilities ${\cal P}_{ij}$, although for recent progress towards this direction see \cite{Urrestilla:2007yw}.  However, we may expect this dependence to be similar to the self-interaction case if these cross-interactions are initiated at a point (at which the incoming strings first cross) and then proceed by the zipping of the colliding strings, rather than exchange of partners. In that case, we may expect the effect of small-scale-structure on the strings to enhance the interaction probability by increasing the number of encounters within one crossing time, much like in the case of self-interactions\footnote{In an alternative picture, in which zipping would require the alignment of the incoming strings along a significant part of their correlation length, we would likely have a different dependence on small-scale-structure than in these point-like interactions.}.  We therefore define:
\be\label{dijk}
d_{ij}^k = d_{ij} S_{ij}^k \, , 
\ee
with                       
\be\label{dij}
d_{ij}= \kappa \times {\cal P}_{ij}^{1/3} \, , 
\ee
where $\kappa$ is a constant of order unity (we will set it to 1).  Note that the effect of a $\kappa \leq 1$ will still be captured to some extent\footnote{Since $w\in (0,1]$ it cannot account for $\kappa \gtrsim 1$.} by our subsequent analysis, because this parameter is degenerate with $w$ (refer to table~\ref{TableFVs} and Eq.~(\ref{PisFV})), which we will vary as an external parameter.  The scaling (\ref{dij}) is another model-dependent choice. Its effect will be examined in section \ref{scaling_gs}. 

The final ingredient that enters the systematic computation of the cross-interaction coefficients $d_{ij}^k$ is the factor $S_{ij}^k$ in Eq.~(\ref{dijk}).  This describes the conditional probability that the crossing of strings $i$ and $j$ produces a zipper of type $k$, given that strings $i$ and $j$ have interacted. The interaction is fully described by an additive and a subtractive channel, so, for each pair $\{i,j\}\equiv \{(p_i,q_i),(p_j,q_j)\}$ there are only two possibilities for $k$; either $k=(p_i+p_j,q_i+q_j)$ or $k=(p_i-p_j,q_i-q_j)$.  Which channel is followed is determined by energetic considerations based on the balance of string tensions at the 3-string junction.  
These are nothing other than the kinematic constraints studied in detail in references \cite{CopKibSteer1,CopKibSteer2,CopFirKibSteer,Salmi}, and which must be satisfied for the junction to form in the first place. Reference \cite{Avgoustidis:2009ke} showed how these microphysical constraints can be integrated over a distribution of velocities and angles in a string network to obtain the averaged network coefficients\footnote{Notice that we have corrected an error in reference \cite{Avgoustidis:2009ke}, where the kinematic constraints were effectively 
overcounted due to the appearance of an extra probability $P_{ij}^k$, see Eqs. (4) and (18) in \cite{Avgoustidis:2009ke}.}:
\be\label{Sijk}
S_{ij}^k = \frac{1}{\cal S} \int_0^1 \int_0^{\pi/2}
\Theta(-f_{\vec\mu}(v,\theta)) \exp[{(v-\bar v_{ij})^2/\sigma_v^{\,2}}]
\, \sin(\theta)\, v^2 {\rm d}\theta {\rm d}v \,,
\ee  
where $\Theta(-f_{\vec\mu}(v,\theta))$ is a step function imposing the 
kinematic constraints $f_{\vec\mu}(v,\theta)<0$ (given by Eq.~(51) of \cite{CopFirKibSteer}), $\sigma_v^2$
is the variance of the velocity distribution, which is assumed to be a Gaussian peaked on the 
relative scaling velocities $\bar v_{ij}=(v_i^2+v_j^2)^{1/2}$, and ${\cal S}$
is a normalisation factor (for details see Eq.~(16) of \cite{Avgoustidis:2009ke}).  Note the direct dependence of these coefficients on the 
string tensions $\mu_i,\mu_j,\mu_k$, and thus on the string coupling $g_s$.  By comparison, 
the dependence on $w$ is very weak and enters only indirectly through the scaling
velocities in $\bar v_{ij}$. 

In table~\ref{TableSs} we show these suppression coefficients for the three lightest string
components (F$\equiv\! 1$, D$\equiv\! 2$, FD$\equiv\! 3$) and for different values of the string coupling $g_s$ 
in the radiation era\footnote{There is a weak dependence of the coefficients $S_{ij}^k$ on the expansion law 
(matter vs radiation era) due to their implicit dependence on the scaling string velocities through the Gaussian 
distribution in Eq. (\ref{Sijk}).}. Due to the weak dependence of these coefficients on $w$, there is no significant 
difference between their values at $w=0.1$ and $w=1$.
\begin{table}[h]
    \begin{center}
    \begin{tabular}{|c|c||c|c|c|} \hline
     $w$ &     $g_s$      & $S_{12}^3$ & $S_{13}^2$ & $S_{23}^1$ \\ \hline\hline  
         & \ $g_s=0.04$ \ &  \ 0.180  \  &  \ 0.293  \   & \  0.950 \    \\
         & \ $g_s=0.1$ \ &  \ 0.117  \  &  \ 0.302  \   & \  0.881 \    \\
         & \ $g_s=0.2$ \ &  \ 0.071  \  &  \ 0.312  \   & \  0.790 \    \\
       $w \in [0.1,1]$  & \ $g_s=0.3$ \ &  \ 0.050  \  &  \ 0.325  \   & \  0.707 \    \\
         & \ $g_s=0.5$ \ &  \ 0.033  \  &  \ 0.354  \   & \  0.590 \    \\
         & \ $g_s=0.7$ \ &  \ 0.028  \  &  \ 0.388  \   & \  0.516 \    \\
         & \ $g_s=0.9$ \ &  \ 0.026  \  &  \ 0.424  \   & \  0.462 \    \\  \hline 
    \end{tabular} 
\end{center}
\caption{\label{TableSs} The coefficients $S_{ij}^k$ for the three lightest string components (F$\equiv\! 1$, D$\equiv\! 2$, FD$\equiv\! 3$) for different values of the free parameters $g_s$ and $0.1 \lesssim w \lesssim 1$, in the radiation era.}
\end{table}

Putting all these factors together, the resulting coefficients $c_i$ and $d_{ij}^k$, in the radiation era, for interactions between the lightest string components are shown in table~\ref{Tablecisdijks} for the same  values of $g_s$ and $w$ as used in table~\ref{TableSs}. In the matter era, the coefficients $c_i$ are a factor $0.78$ smaller, while the $d_{ij}^k$ coefficients change only through the implicit velocity dependence in Eq.~(\ref{Sijk}).  This change in $d_{ij}^k$ does not affect significantly the scaling values of the string correlation lengths and velocities, and so can be neglected. The main difference with reference \cite{Avgoustidis:2009ke}  is that in \cite{Avgoustidis:2009ke} the coefficients $d_{ij}$ were chosen independently as free parameters, while now they are systematically computed as described above and they depend only on $g_s$ and $w$, which are our free parameters.     
\begin{table}[h]
    \begin{center}
    \begin{tabular}{|c|c||c|c|c|c|c|c|} \hline
     $w$ &     $g_s$      & $c_1$ & $c_2$ & $c_3$ & $d_{12}^3$ & $d_{13}^2$ & $d_{23}^1$ \\ \hline\hline  
         & \ $g_s=0.04$ \ & 0.02 & 0.13 & 0.13 & 0.05 & 0.08 & 0.55 \\
         & \ $g_s=0.1$ \ & 0.03 & 0.16 & 0.16 & 0.04 & 0.11 & 0.62  \\
         & \ $g_s=0.2$ \ & 0.05 & 0.19 & 0.19 &  0.03 & 0.14 & 0.63  \\
     $w=1$ & \ $g_s=0.3$ \ & 0.07 & 0.20 & 0.20 & 0.03 & 0.16 & 0.61  \\
         & \ $g_s=0.5$ \ & 0.10 & 0.21 & 0.21 & 0.02 & 0.21 & 0.54 \\
         & \ $g_s=0.7$ \ & 0.12 & 0.22 & 0.22 & 0.02 & 0.26 & 0.49 \\
         & \ $g_s=0.9$ \ & 0.15 & 0.22 & 0.22 & 0.02 & 0.31 & 0.45 \\  \hline
\end{tabular} \ \ \ \ 
\begin{tabular}{|c|c||c|c|c|c|c|c|} \hline
     $w$ &     $g_s$      & $c_1$ & $c_2$ & $c_3$ & $d_{12}^3$ & $d_{13}^2$ & $d_{23}^1$ \\ \hline\hline  
          & \ $g_s=0.04$ \ & 0.01 & 0.13 & 0.13 & 0.05 & 0.07 & 0.55 \\
          & \ $g_s=0.1$ \ & 0.02 & 0.16 & 0.16 & 0.04 & 0.10 & 0.62 \\
          & \ $g_s=0.2$ \ & 0.02 & 0.19 & 0.19 & 0.03 & 0.13 & 0.63 \\
       $w=0.1$ & \ $g_s=0.3$ \ & 0.03 & 0.20 & 0.20 & 0.02 & 0.14 & 0.61 \\
          & \ $g_s=0.5$ \ & 0.05 & 0.20 & 0.21 & 0.01 & 0.15 & 0.54 \\
          & \ $g_s=0.7$ \ & 0.06 & 0.15 & 0.22 & 0.01 & 0.17 & 0.39 \\
          & \ $g_s=0.9$ \ & 0.07 & 0.12 & 0.21 & 0.01 & 0.20 & 0.31 \\  \hline   
    \end{tabular}   
    \end{center}
\caption{\label{Tablecisdijks} The coefficients $c_i$ (in radiation era) and $d_{ij}^k$ of Eq.~(\ref{rho_idtgen}-\ref{v_idtgen}) for the three lightest string components (F$\equiv\! 1$, D$\equiv\! 2$, FD$\equiv\! 3$) for different values of the free parameters $g_s$ and $w$. The values of $c_i$ in the matter era are obtained by multiplying the $c_i$ in this Table by a factor ${\tilde c}_m / {\tilde c}_r = 0.78$ (see Eq.~(\ref{cis})).  Finally, the values of $d_{ij}^k$ in the matter era change only through the implicit velocity dependence in Eq.~(\ref{Sijk}) and this change does not affect significantly the scaling values of the string correlation lengths and velocities.}  
\end{table}          

\subsubsection{Average length of zippers}

The average length $\ell_{ij}^k(t)$ of zippers produced at time $t$ which appears in Eq.~(\ref{rho_idtgen}-\ref{v_idtgen}) is, in principle, model-dependent.  However, for cosmic superstrings, which are not subject to topological conditions which exist for example in non-Abelian field theory strings, it can be taken to be \cite{NAVOS,Avgoustidis:2009ke}
\be 
\ell_{ij}^k=\frac{L_i L_j}{L_i+ L_j} \,. 
\label{zipper-length}
\ee
With this choice we are assuming that the produced zipper has a length which is smaller than -- but close to -- the smallest of the two correlation lengths of the colliding strings. 

\subsubsection{The $b_{ij}^k$ coefficients in the velocity Eq.~(\ref{v_idtgen})}

For simplicity we set the coefficients $b_{ij}^k$ to zero. In this way we 
concentrate on the effects coming from the junction terms in Eq.~(\ref{rho_idtgen}) and not 
from the extra terms in the velocity evolution Eq.~(\ref{v_idtgen}). In section \ref{scaling_gs} we will also discuss the case with $b_{ij}^k=d_{ij}^k$ (refer 
also to section \ref{multimu}).  In agreement with \cite{Avgoustidis:2009ke} 
we find that our key results are insensitive to this choice. 

\subsubsection{Radiation to matter era transition}

Finally, when solving Eqs.~(\ref{rho_idtgen}-\ref{v_idtgen}) we must of course run through the regimes of radiation domination to matter domination and finally to a $\Lambda$ dominated era. Of particular importance is the transition from radiation to matter domination and we interpolate between these two eras by following the approach in 
\cite{Battye:1997hu,Pogosian:1999np}, namely by replacing, for each string type,
\be
\label{tilde-c-rad-mat}
c_i = \frac{\tilde{c}_r+ga \tilde{c}_m}{1+ga} \times \mathcal{P}_i^{1/3}.
\ee 
Here $\tilde{c}_r=0.23$ (radiation), $\tilde{c}_m=0.18$ (matter), $g=300$, and the scale factor $a(t)$,  calculated numerically from the Friedmann equation, is normalised so that $a=1$ today. 

\subsection{Scaling of cosmic superstring networks at large and small string couplings}\label{scaling_gs}

We have solved Eq.~(\ref{rho_idtgen}-\ref{v_idtgen}) numerically using the parameters computed in section \ref{superparams} for different values of our external parameters $g_s$ and $w$.  As in references \cite{TWW,NAVOS,Avgoustidis:2009ke}, we find scaling solutions in which all network components reach approximately constant string number densities and rms velocities during the radiation and matter eras. 

The string number density per unit Hubble volume is defined by 
\be
\xi_i^{-2} \equiv (t/L_i)^2.
\label{str-no}
\ee 
We find that it is dominated by the lightest three network components, namely the F, D and FD strings, while all heavier components generally have negligible number densities. The same three strings also dominate the effect on the CMB observables or, more generally, any observable linearly related to the two point function of the string energy momentum tensor. With that in mind, it is useful to introduce the {\it power spectrum density} $M_i$ given by \cite{PogosianTye}
\be\label{Mdens}
M_i = \left(\mu_i \over \xi_i \right)^2 \ ,
\ee
which determines the amplitude of string induced power spectra.

\begin{figure}[tbp]
\begin{center}
\includegraphics[scale=0.5]{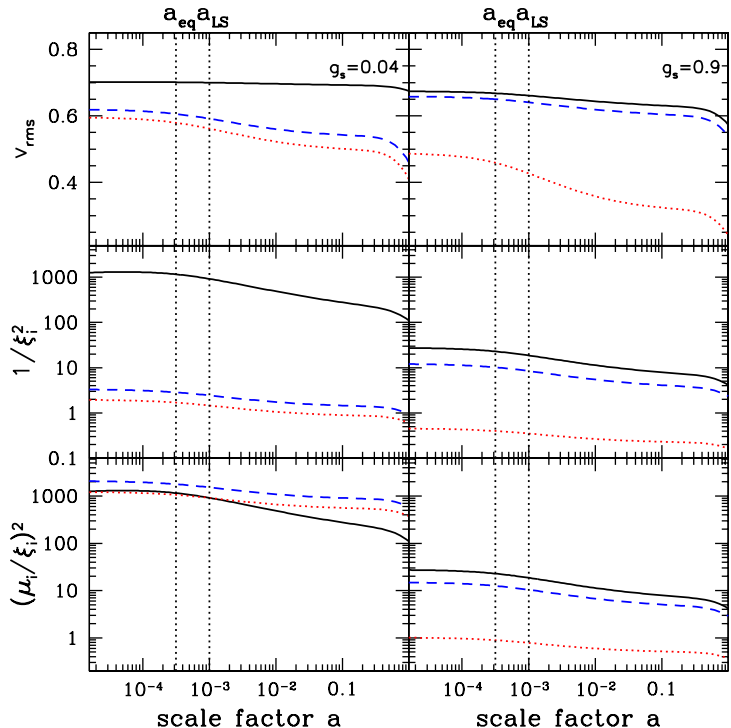}
\caption{\label{fig1} Evolution of the rms velocity $v_i$ (top panels), number density $\xi_i^{-2} \equiv (t/L_i)^2$ (middle panels), and the   power spectrum density $(\mu_i/\xi_i)^2$ (bottom panels) of the three lightest network components: F strings (solid black), D strings (blue dash), and FD strings (red dot), at two representative values of the string coupling $g_s$. The panels in the left column are for $g_s=0.04$, those on the right are for $g_s=0.9$. All plots are for $w=1$. For $g_s \rightarrow 1$, the tensions of the F and D strings are comparable, as well as their densities. At smaller $g_s$, the lighter F strings dominate the {\it number} density, while the heavier and less numerous D strings dominate the power spectrum. The epochs of radiation-matter equality and last scattering are indicated with vertical lines. Of particular relevance for CMB is the correlation length of the string type that dominates the power spectrum density at the time of LS (see also Figs.~\ref{fig2} and \ref{fig3}). The deviation from scaling at late times is simply due to the fact we are entering a $\Lambda$ dominated era.}
  \end{center}
\end{figure}

Fig.~\ref{fig1} shows the evolution of the rms velocity, number density and the power spectrum density for the three lightest strings for $g_s=0.04$ and $g_s=0.9$, with $w=1$. When $g_s$ is close to unity, the tensions of the F and D strings are comparable. As a result, their densities are comparable too, as shown in the panels on the right. When  $g_s$ is decreased, the lighter F strings become more populous and thus dominate the {\it number} density of the network. However, the D strings become heavier at smaller $g_s$ and, while less numerous, can dominate the power spectrum density, as evident from the $g_s=0.04$ case shown in the left panels of Fig.~\ref{fig1}. This transition from the power spectrum being dominated by the light populous F strings at higher string couplings to heavy rare D strings at smaller couplings appears to be a generic property of cosmic superstring networks and is one of the key results of this paper. Reducing the volume parameter $w$ to 0.1 results in an enhancement of the overall network number density (since ${\cal P}_{ij}\propto {\cal V}_{ij}\propto w$), but the dependence of the network scaling patterns on $g_s$ remains the same.  

\begin{figure}[tbp]
\begin{center}
\includegraphics[scale=0.5]{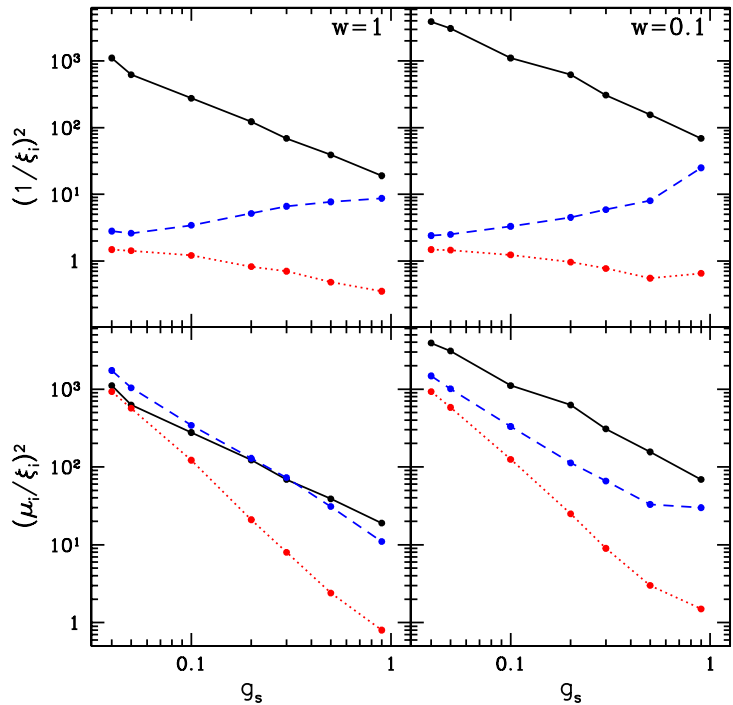}
\caption{\label{fig2} Dependence of the string number (top panels) and power spectrum (bottom panels) densities {\it at the time of last scattering} on the value of the string coupling $g_s$ for $w=1$ (left) and $w=0.1$ (right), for the three lightest network components: F strings (solid black), D strings (blue dash) and FD strings (red dot).}
\end{center}
\end{figure}

In Fig.~\ref{fig2}, we plot the number and power spectrum densities of the three relevant strings at the time of last scattering vs $g_s$ for two values of $w$. Reducing the value of the string coupling from $g_s\approx 1$, initially reduces the contribution of D strings to $M_{\rm total}$ as they become more rare, but this trend quickly changes as we continue reducing $g_s$ and the D strings become heavy $\mu_{\rm D}\simeq g_s^{-1}$, eventually dominating $M_{\rm total}$ in the case of $w=1$. For $w=0.1$, the power density $M_D$ increases with decreasing $g_s$, but does not quite catch up with $M_F$ over the range of string couplings we have studied numerically. Note, however, that the number density of F becomes a few orders of magnitude larger that that of D, meaning that the correlation length of F strings is very small. If their correlation length is smaller than the horizon size at last scattering, the F strings do not contribute significantly to the CMB polarisation and, despite their lower power density, the $B$ mode spectrum will in fact be dominated by D strings. This will be discussed more in the next section.
     
Having sketched this interesting trend in the dependence of network scaling patterns on $g_s$, let us see the way it emerges from the functional forms for the MTSN parameters adopted in Section~\ref{superparams}. At {\it large} string couplings ($g_s \rightarrow 1$), the hierarchy in the string tensions is not pronounced ($\mu_{\rm D}=g_s^{-1}\mu_{\rm F}\sim \mu_{\rm F}$), and the F and D strings can be comparable both in number and power spectrum densities. In particular, for $w\sim 1$ and $g_s\lesssim 1$, we have ${\cal P}_{\rm F}\sim g_s^2 w \lesssim {\cal P}_{\rm D}\sim 1$, so, in effect, one would observe a single effective network of tension $\mu_{\rm F}\sim \mu_{\rm D}$, with a correlation length similar to that of ordinary cosmic strings, but with the extra property of frequently forming 3-string junctions.  For small $w$, the situation would be similar, but with a smaller correlation length, and the appearance of 3-junctions would be more rare since now ${\cal P}_{\rm F}\sim g_s^2 w\lesssim w \ll 1$ is somewhat smaller than ${\cal P}_{\rm D}\sim w g_s^{-3} > w$, so D strings would be somewhat more rare. It should be noted, however, that at $g_s\approx 1$ the perturbative methods used to calculate ${\cal V}_{\rm FF}$ are not expected to be accurate.

For {\it small} $g_s$, the difference between the tensions of F and D strings is large, and so is the difference in the corresponding coefficients $c_i,d_{ij}^k$ (see table~\ref{Tablecisdijks}). As a result, the much lighter F strings dominate the network number density. The power spectrum density, however, is dominated in this case by the less populous, but much heavier ($\mu_{\rm D}\propto g_s^{-1}\mu_{\rm F}$) D strings which, being very massive, evolve practically independently of the light F-string network. Indeed, even the zipping between F and D strings gives rise to FD composites with a tension practically equal to that of the heavy D strings. So, as far as the contribution to the power spectrum is concerned, one has effectively a single string network of D strings -- the F string component remains unobservable due to its low tension, despite the fact it has a high number density. Moreover, unless $w\ll g_s^3$, the volume factors ${\cal V}_{\rm DD}\propto wg_s^{-3}$ in Eq.~(\ref{PisFV}) approach unity while the non-perturbative factor ${\cal F}_{\rm DD}\lesssim 1$. As a result, the network properties are similar to those of an ordinary\footnote{Note that for extremely small $g_s$ it is possible that this behaviour changes again to one with $c_{\rm D}\ll 1$ as the non-perturbative amplitude is expected to fall to ${\cal F}_{\rm DD}\ll 1$ as $g_s\rightarrow 0$ (see \cite{Jackson:2004zg}).  This effect, however, is not understood quantitatively at present.} field theory string network with $\tilde c\approx {\cal P}\approx 1$. For $w\ll g_s^3$ the situation is similar, but the number density is enhanced (smaller correlation length) compared to ordinary cosmic strings.

There may also be an intermediate regime, in which the string coupling $g_s$ is small enough for the light F strings to dominate the number density, but large enough for the heavier D strings to contribute significantly to the power spectrum density without dominating it completely. In this case there is a notable difference in the coefficients $c_F$ and $c_D$ and there are two substantially different correlation lengths describing the network.  Any possible observational effect from this hybrid network would depend on some combination of the two correlation lengths that could be significantly different than for ordinary cosmic strings. Such an intermediate regime, while potentially interesting, is not the main focus of this paper, and we will leave discussion of it for future work.           

\begin{figure}[h]
\begin{center}
\includegraphics[scale=0.5]{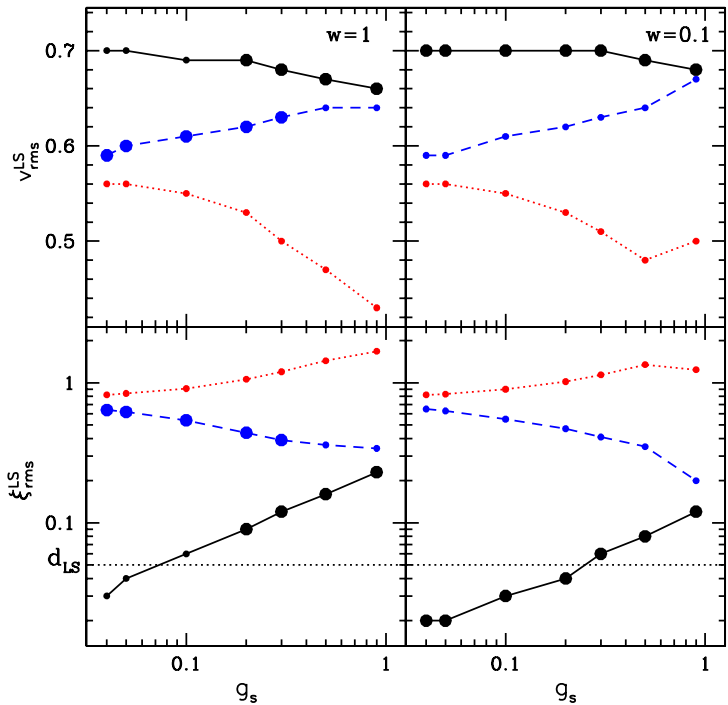}
\caption{\label{fig3} 
The correlation length and the rms velocity \textit{at the time of last scattering} (LS) as a function of the string coupling $g_s$ for $w=1$ (left) and $w=0.1$ (right) for the three lightest network components: F strings (solid black), D strings (blue dash) and FD strings (red dot). We show the string type(s) that dominate(s) the power spectrum at LS with oversized points. The horizontal line at $\xi=0.05$ indicates the thickness of the LS surface.}
\end{center}
\end{figure}

In the next section we study possible observable signals induced by these networks on the CMBR temperature and polarisation power spectra. Of particular relevance in this regard are the values of the correlation length ($\xi_{\rm LS} = L(t_{\rm LS})/t_{\rm LS}$) and the rms velocity ($v_{\rm LS}$) of the string type that dominates the power spectrum density at the time of last scattering (LS). In Fig.~\ref{fig3} we show the dependence of $\xi_{\rm LS}$ and $v_{\rm LS}$ on $g_s$. The fact that strings of different correlation lengths dominate the spectra at different values of $g_s$ means that there could well be distinctive signals in the CMB spectra as a function of $g_s$. This may allow us to use CMB spectra as a discriminant between these different types of limiting behaviour, potentially providing an invaluable tool for constraining the underlying string theory, in particular the string coupling $g_s$ and, more indirectly, the scale of compactification. We also note that, as seen in Fig.~\ref{fig3}, there is only a minor difference in the values of scaling velocities of the dominant string types at different couplings. 

Before ending this section let us briefly discuss the dependence of our results on the various model-dependent  
assumptions and approximations we have adopted. In agreement with 
\cite{Avgoustidis:2009ke} we find that our results are practically identical when 
$b_{ij}^k=0$ and $b_{ij}^k=d_{ij}^k$. We have also considered deviations from the 
functional form of the volume factors ${\cal V}_{ij}$ for $(p,q)$-strings (third and fifth 
entries of Table~\ref{TableFVs}, right). Since the networks are dominated by the lightest 
strings, such deviations only affect the heavy, rare species, so our scaling results
are robust at the percent level, except if $g_s\simeq 1$. An interesting 
model-dependent extension of our string evolution model is to allow for 
spontaneous decay (unzipping) of heavy strings into lighter ones, by adding 
the appropriate new terms in Eq.~(\ref{rho_idtgen}). The effect of these 
terms is to transfer length from heavy to light strings, but, 
again, since our networks are dominated mostly by F and D strings, with heavier
composites being suppressed, these terms only give rise to corrections at percentage
level in most of the parameter space we consider\footnote{When the number density of FD strings 
is comparable to that of D strings, then spontaneous decay leads to a reduction of the FD string
density in favour of D and F strings, and this effect is important, in agreement with the findings 
of Ref. \cite{Urrestilla:2007yw}.  However, for cosmic superstrings, the combination of ${\cal P}_{ij}$
and $S_{ij}^k$ (refer to Eqs. (\ref{PisFV}) and (\ref{dijk}-\ref{Sijk})) generally implies that the FD string
density is smaller than that of D strings (top panel of Fig. \ref{fig2}) and this effect is negligible. The effect 
is strongest for small $g_s$, when the densities of FD and D strings are closer, but in this case 
the (number) density of F strings is higher by at least two orders of magnitude.  Thus, the effect of 
spontaneous unzipping in this case, is at most an order unity enhancement of the D string density 
and our results (which depend on the relative density and tension between F and D strings) remain 
robust.}. The most important model-dependent 
assumptions are the scaling of $d_{ij}$ with ${\cal P}_{ij}$, Eq.~(\ref{dij}), and the 
adoption of a non-perturbative amplitude ${\cal F}_{ij}$ for interactions containing 
D strings in both of the colliding segments (fourth and fifth entries in Table~\ref{TableFVs}, left). Even though the two limiting behaviours discussed above are qualitatively robust, 
the transition point between them, as well as the values of the correlation length of the 
dominant strings in each regime, depend quantitatively on both of these choices.  This highlights the need for investigating numerical simulations of strings with junctions and for further developing non-perturbative techniques for D string interactions. 

Finally, we comment on the significance of using the one-scale assumption to model the evolution of the network. To test the role of this assumption in setting the hierarchy of densities of different string types we have incorporated a second scale in our model, following the approach of \cite{NAVOS}, i.e. adding an extra scale $\bar{\xi}$, common to all string species, and modifying appropriately the evolution equations for the correlation length and the velocity. We have concluded that the effect of the second scale is not large -- the number densities of the strings do not significantly change and, most importantly, the hierarchy stays the same. Another potential concern, which is of relevance to the CMB predictions, is that the one-scale assumption equates the correlation length along the strings with the interstring distance. This is only a reasonable approximation for networks with sufficiently large intercommutation probabilities (${\cal P}_i \gtrsim 0.1$). Hence, we do not consider values of $w$ and $g_s$ at which intercommutation probabilities of the dominant strings are too small.

\section{CMB temperature and B-mode spectra from multi-tension strings}
\label{sec:cmb}

Having obtained the MTSN scaling solutions at different values of string couplings, we would like to examine their imprint on the CMB temperature and polarisation spectra. First, we describe how we calculate CMB spectra for MTSN, then turn our attention to the shapes of the spectra at large and small values of $g_s$.

\subsection{Modeling CMB with CMBACT}
\label{sec:cmbact}

To evaluate the CMB temperature and polarisation spectra sourced by multi-tension string networks (MTSN) we modify the publicly available code CMBACT \cite{Pogosian:1999np,cmbact} so as to allow for strings with multiple tensions whose scaling is modelled by Eqs.~(\ref{rho_idtgen}) and (\ref{v_idtgen}). 

In CMBACT, the string network is represented as a collection of uncorrelated string segments, an approximation proposed in \cite{Vincent:1996qr} and adapted for calculation of CMB spectra in \cite{Albrecht:1997nt,ABR99,Pogosian:1999np}. In the unconnected segment model (USM), straight segments of strings are produced at some early time and given random/uncorrelated orientations and velocities. At later times, a certain fraction of the number of segments decays in such a way as to match the number density given by a scaling model. The initial  positions and orientations of the segments are drawn from uniform distributions, and the direction of the velocity is taken to be uniformly distributed in the plane perpendicular to the string orientation (longitudinal velocities are neglected). 

In the default version of CMBACT, the key parameters of the segments --- namely their length, rms velocity and number density --- are modelled using the VOS Eq.~(\ref{vosrho}) and (\ref{vosvel}). The USM does not explicitly follow the loop distribution, however the energy in the loops is effectively included as part of the covariant conservation of the energy momentum of the scaling network. On their own, the straight string segments with open ends
violate energy conservation. To remedy this, CMBACT enforces energy conservation by calculating the components $T_{00}$ and $T_{ij}$ (with $i \ne j$) of the energy momentum tensor and then using the covariant conservation equation $\nabla^\mu T_{\mu\nu}=0$ to calculate $T_{0i}$ and $T_{ii}$. 

We emphasise that CMBACT is not a means for gaining new insight into the evolution of cosmic string networks. Instead, it is a tool for evaluating CMB spectra for {\it given} one-scale parameters, such as the correlation length and rms velocity. In \cite{Battye:2010xz}, it was shown that the CMB spectra obtained from field theoretical simulations of Abelian-Higgs (AH) strings \cite{Bevis:2006mj} are reproduced by CMBACT when the one-scale parameters measured in the simulation are used as input. Also, the CMB spectra obtained from the NG simulations of \cite{Ringeval:2005kr,Fraisse:2007nu} were compared to those from CMBACT with the one-scale parameters measured in \cite{Ringeval:2005kr}, and good agreement was once again obtained. The default version of CMBACT uses the VOS model with parameters tuned to match NG simulations of \cite{Shellard:1989yi,VOSk}.

The shapes of the string-induced CMB spectra are mainly determined by the large-scale properties of the string network, such as the correlation length and rms velocity. The overall normalisation of the spectrum has a simple dependence on the string tension $\mu$ and $\xi = L/t$ \cite{PogosianTye,Pogosian:2007gi} given by Eq.~(\ref{Mdens}). The CMB temperature spectrum (TT) receives a contribution from the LS surface, for which the relevant scale is $\xi$ at the time of LS. In addition, TT receives roughly equal contributions at each subsequent epoch (which is the mechanism by which strings can produce a scale-invariant TT spectrum on large scales) and hence the value of $\xi$ is approximately the value measured during matter domination. CMB polarisation, on the other hand, is sourced at the time of LS and thus the normalisation of the B-mode spectrum is given by $\xi_{\rm LS}$.

In this work we generalise CMBACT to include unconnected segments of $N$ different types. The lengths and rms velocities of each type are determined from Eqs.~(\ref{rho_idtgen}) and (\ref{v_idtgen}). As in the single-tension case, the $T_{0i}$ and $T_{ii}$ components of the string stress-energy are determined from the covariant conservation equation, which now takes the form
\be
\nabla^\mu \displaystyle\sum_{i=1}^{N} T_{\mu\nu}^{i}=0.
\ee
The overall amplitude of the CMB angular spectra $C_\ell$ is approximately determined by
\be
C_\ell^{strings} \propto M_{\rm total} = \sum_{i=1}^{N} M_i = \sum_{i=1}^{N} \left(\frac{\mu_i}{\xi_i}\right)^2,
\label{clstringN}
\ee
and the shapes of the spectra will be set by the correlation length and the rms velocity of the most dominant population
of strings.

We should note that CMBACT has only been tested against simulations of single tension strings with no junctions
and with an intercommutation probability of unity. However, as discussed in Sec.~\ref{sec:scaling}, the scaling solution of our FD networks tends to fall into two categories depending on whether the string coupling is large or small. Namely, the energy density of the network is dominated by light populous strings for large values of $g_s$, and rare heavy strings at much smaller $g_s$. In either case, the junctions are rare, and the bulk of the anisotropy is seeded by a single type of strings. Moreover, the intercommutation probabilities of the dominant string species are $0.1$ or larger implying at most a factor of $2$ reduction in loop chopping efficiency $c_i$. This justifies the use of CMBACT for modelling the CMB spectra from FD networks at least in the two limiting cases of large and small string couplings.

\subsection{The CMB spectra for scaling FD strings}
\label{sec:cmb-results}

Before proceeding to discuss the CMB spectra sourced by FD strings, let us first note that cosmic strings cannot contribute more than $10\%$ of the total CMB temperature anisotropy \cite{contaldi, battye, bouchet, bevis,Wyman:2005tu, Seljak:2006bg, Bevis:2007gh,Battye:2010xz}. To comply with this bound, we will adjust the fundamental (F) string tension $\mu_F$ to be such that 
\be
f_s=C^{TT}_{strings} / C^{TT}_{total} = 0.1 \ ,
\ee
where we follow conventions of \cite{Pogosian:2007gi} to define
\be
C^{TT} \equiv \sum_{\ell=2}^{2000}(2\ell+1)C_{\ell}^{TT}.
\ee
Fortunately though, even with a marginal contribution to the $TT$ spectrum, strings can be a prominent source of $B$ mode polarisation. This is because strings, unlike inflation, are actively sourcing vector mode perturbations of magnitude comparable to the scalar perturbations (see \cite{Seljak:1997ii, Battye:1998js, PogosianTye, Bevis:2007qz,Pogosian:2007gi,Urrestilla:2008jv} for work on the subject). 

The dependence of the TT and BB power spectra on the correlation lengths and the rms velocities was extensively studied in \cite{Pogosian:2007gi} and more recently in \cite{Battye:2010xz}. The overall amplitude of the spectrum is approximately given by Eq.~(\ref{clstringN}). The correlation length and the rms velocity of the main string type set the dominant momentum modes in the strings stress-energy, which in turn determine the position of the main peak. Larger string correlation lengths will move the peaks in the TT and BB spectra to lower $\ell$. The dependence of the peak position on the rms velocity is non-monotonic. The positions of the TT and BB peaks move to higher multipoles (smaller scales) for low and moderate velocities, but move to larger scales (lower $\ell$) for higher velocities. Also, larger values of $v$ somewhat decrease the amount of BB power relative to TT power. The current bound on the fraction of string sourced CMB temperature anisotropy is only weakly dependent on the detailed shape of the spectrum, which in turn depends on the different types of strings involved.  For instance, the bound on global strings, which do not lead to a pronounced peak in the CMB spectrum, is also approximately $10\%$ \cite{bevis}. 

\begin{figure}[tbp]
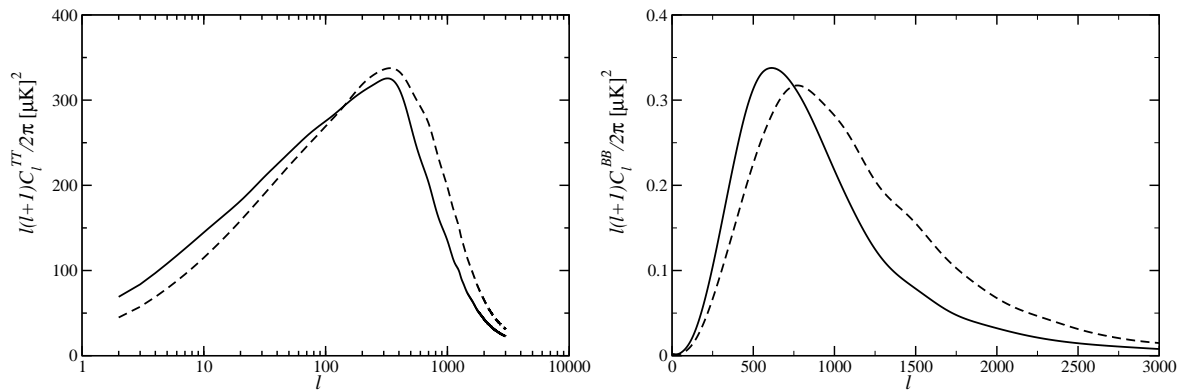

\centering
\includegraphics[scale=0.3]{TTcomp_P13S_w1_gs004_09.eps}
\includegraphics[scale=0.3]{BBcomp_P13S_w1_gs004_09.eps}
\caption{The normalised TT (left) and BB (right) power spectra for $g_s=0.04$ (solid) and $g_s=0.9$ (dash) for $w=1$, normalised to give $f_s=0.1$. Note that the smaller string coupling leads to a discernible move in the peak of the BB spectra to smaller $\ell$.}
\label{TTsuper}
\end{figure}

\begin{figure}[tbp]
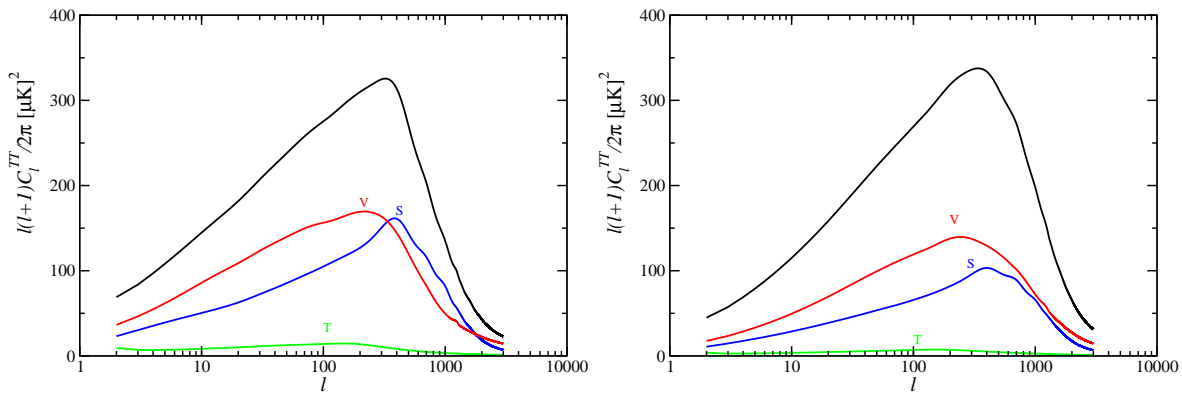

\centering
\includegraphics[scale=0.3]{P13S_gs004SVT_w1.eps}
\includegraphics[scale=0.3]{P13S_gs09SVT_w1.eps}
\caption{The normalised TT power spectra for $g_s=0.04$ (left) and $g_s=0.9$ (right), including the individual scalar (S), vector (V) and tensor (T) contributions.}
\label{TTSVT}
\end{figure}

In Fig.~\ref{TTsuper} we show the TT and BB power spectra for two values of the string coupling $g_s$ (solid black line for $g_s=0.04$, and dashed line for $g_s=0.9$). In Fig.~\ref{TTSVT} we show the normalised total TT power spectra for $g_s=0.04$ (left) and $g_s=0.9$ (right), including the individual scalar (S), vector (V) and tensor (T) contributions for each case. The scalar and vector contributions are of similar magnitude, as expected in the case of cosmic strings. The spectra are normalised to give $f_s=0.1$ as described earlier, which translates into $G\mu_F=1.8 \cdot 10^{-8}$ for $g_s=0.04$, and $G\mu_F=2.1 \cdot 10^{-7}$ for $g_s=0.9$.

We can interpret the CMB spectra at different values of $g_s$ in the context of the scaling behaviour of the different string types discussed in detail in Sec.~\ref{sec:scaling}. In the case $g_s=0.9$, the F and D strings are dominant and most populous, having almost the same tension. Their correlation lengths and velocities are similar as well, with values close to those of ordinary strings. Hence, their contributions to the CMB spectra are comparable, whereas the contribution from the heavier FD string is not as important. On the other hand, in the $g_s=0.04$ case, the F string is again light and populous, but the D string is $25$ times heavier than the F string, and so is the FD string. Thus, despite being very rare, the D and FD strings dominate the CMB spectra. In addition, there is a small but non-negligible contribution from the heavier $(2,1)$ string.

The right panel in Fig.~\ref{TTsuper} clearly shows the impact of the changed hierarchy between the three kinds of strings. The heavy D strings which dominate $M_i$ for $g_s=0.04$ have a larger correlation length, which translates into a BB peak at smaller $\ell$. This offers a tantalising possibility for probing for small string couplings in the CMB.

\begin{figure}[tbp]
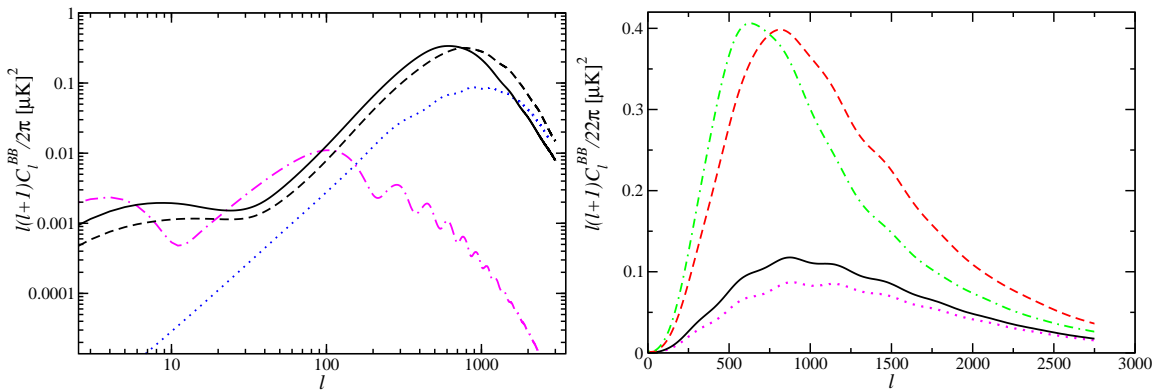

\centering
\includegraphics[scale=0.3]{BBprimlenscomp.eps}
\includegraphics[scale=0.3]{lens_gs004_09.eps}
\caption{{\it Left:} The B-type polarisation spectra due to cosmic superstrings assuming a $10\%$ contribution ($f_s=0.1$) are plotted with solid ($g_s =0.04$) and dashed ($g_s =0.9$) 
black lines. The expected $C_l^{BB}$ spectra for E to B lensing (blue dot line) and from primordial gravitational waves assuming a tensor-to-scalar ratio of $r=0.1$ (magenta-dot-dash line) are shown for comparison. {\it Right:} The magenta dot line is the lensing prediction, the black solid line is the sum of the string and lens-sourced B-mode power for $g_s=0.9$ for $f_s=0.01$. Strings manifest themselves via the systematic excess power at high-$\ell$ over the lensing prediction. The sum of strings and lensing contributions is also plotted for $f_s=0.1$ for $g_s=0.9$ (red dash) and $g_s =0.04$ (green dot-dash). By measuring the location of the main peak we can rule out either the small or the large values of $g_s$.}
\label{BBprimlens}
\end{figure}

\begin{figure}[tbp]
\centering
\includegraphics[scale=0.35]{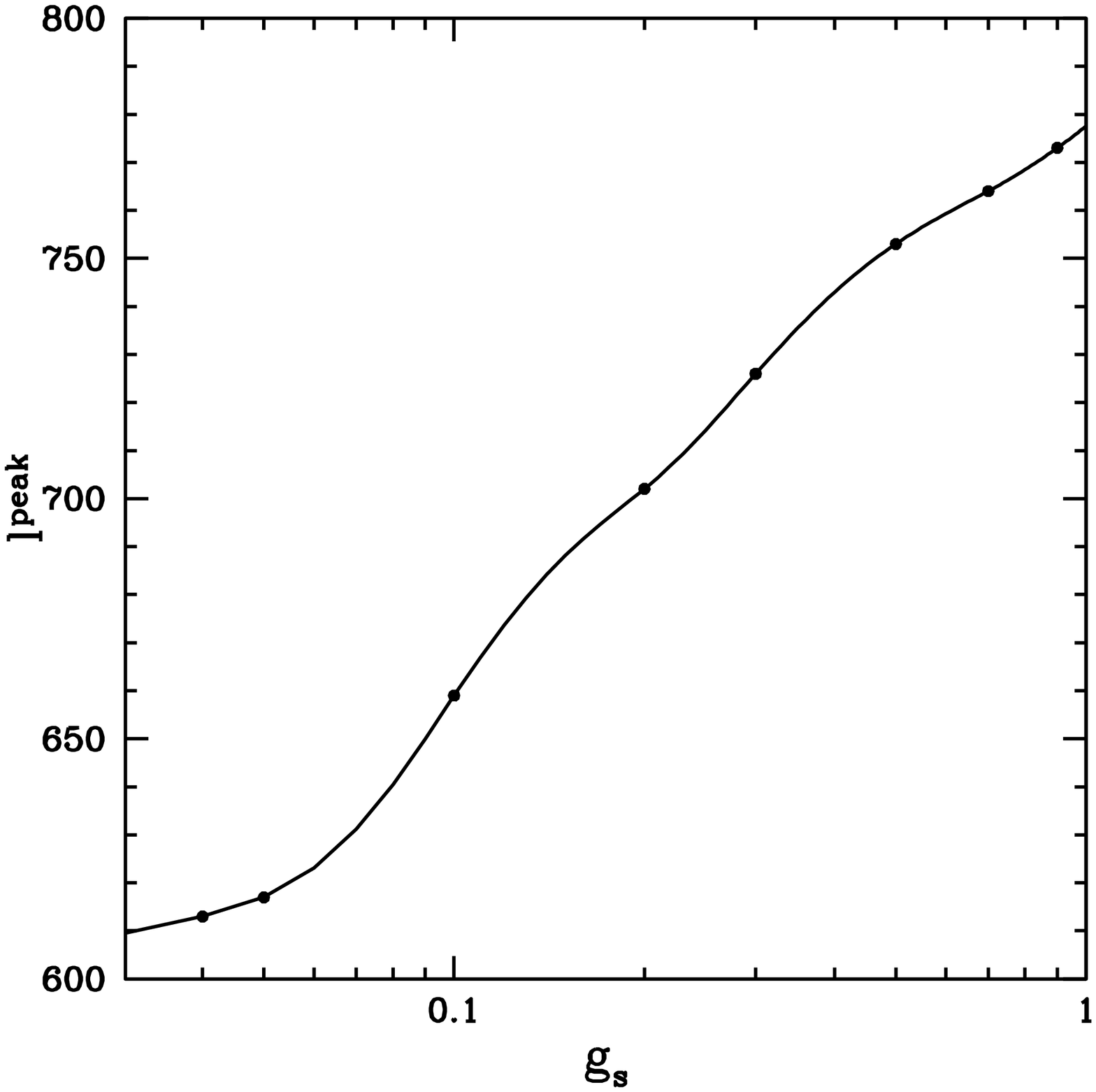}
\includegraphics[scale=0.35]{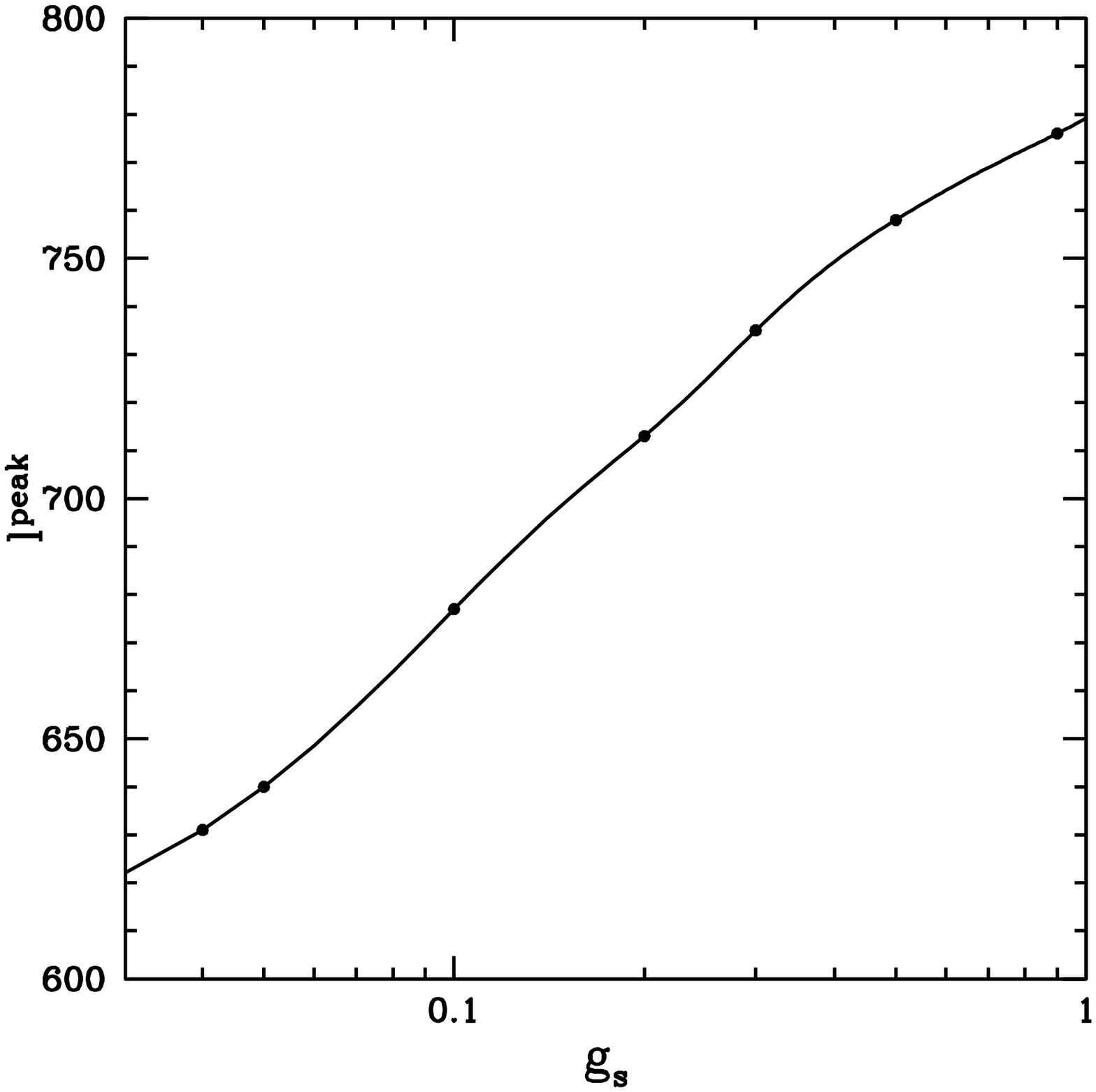}
\caption{The position of the peak of the BB spectrum as a function of the string coupling $g_s$ for $w=1$ (left) and $w=0.1$ (right).}
\label{BBpeakw1}
\end{figure}

In Fig.~\ref{BBprimlens} we again show the B-type polarisation spectrum predicted by our string models for a $10\%$ and a $1\%$ contribution to the total TT. We compare this to the contributions from gravitational lensing of the adiabatic E-mode polarisation into B-modes. Of note, especially in the region of high $\ell$, is the existence of a possible detection window, where the cosmic string signal would manifest itself as an excess over the expected lensing contribution. The Planck satellite may be able to see the excess if strings contribute at a level currently tolerated by data \cite{Bevis:2010gj}, however, it is likely that Planck's TT, TE, and EE spectra will place tighter bounds on strings even without B-mode measurements. The science goals of the ground based experiments, such  as the advanced stages of QUIET \cite{quiet} and POLARBEAR \cite{polarbear}, include accurate measurements of the BB spectrum from lensing. If these science goals are met, they should have the sensitivity to detect the excess due to strings at a level of $f_s \sim 10^{-3}$. If the string contribution is sufficiently large, it may be possible to detect the main peak and thus rule out large or small values of $g_s$ based on the position of the peak. More studies are needed to determine the minimum value of $f_s$ for which a particular B-mode experiment will be able to detect the position of peak \cite{Mukherjee:2010ve,MossPogosian}. In principle, it may be possible to de-lense the B-mode polarisation map, taking advantage of the fact that the B-mode due to lensing is a rotation of the E-mode, and hence E and B modes are highly correlated \cite{Okamoto:2003zw,Smith:2008an}. Some preliminary forecasts of expected constraints on cosmic strings from de-lensed B-modes were reported in \cite{SelSlo,Mukherjee:2010ve} but more work is needed to understand the feasibility of measuring the position of the string induced peak from de-lensed polarisation maps \cite{MossPogosian}.

In Fig.~\ref{BBpeakw1} we show the BB peak location as a function of $g_s$, ranging from $0.04$ to $0.9$. For $w=1$, shown on the left plot, we see that the position of the BB spectrum as a function of the string coupling $g_s$ is decreasing with decreasing $g_s$. This is consistent with our earlier discussion for the behaviour of the power spectrum density, which starts being dominated by the F strings and, after a transition, ends up being dominated by the heavy rare D strings. Their correlation length increases with decreasing $g_s$, and the BB peak moves to smaller $\ell$. 

In the right plot of Fig.~\ref{BBpeakw1}, which considers the $w=0.1$ case, we also see that the BB peak position decreases with a decreasing $g_s$. However, here it happens for somewhat different physical reasons. Namely, as we see in Fig.~\ref{fig2}, the power spectrum density $M_i$ in the more populous F strings dominates through the whole range of $g_s$ we have considered. Yet it is again the D string contribution that dominates the B-mode spectra. The explanation for this lays on the smallness of the $\xi_i$ for the F strings. For $\xi \lesssim 0.05$, the string correlation length is smaller than the thickness of the LS surface, and thus most of the power in that type of string does not contribute to the B-mode. Only a small fraction of the total power in F strings contributes, that generated on larger scales, but it is much smaller than the contribution of the D strings. The important issue arising here is that when the $\xi$ of the dominant string becomes so small, the one-scale approximation is no longer reliable. This is a well-known problem, and further improvements for accurately calculating the CMB contributions from such strings must be performed in the future.

\section{Combined constraints on $\mu_F$ and $g_s$ from CMB and Pulsar Timing}
\label{sec:pulsars}

\begin{figure}[tbp]
\includegraphics[scale=0.4]{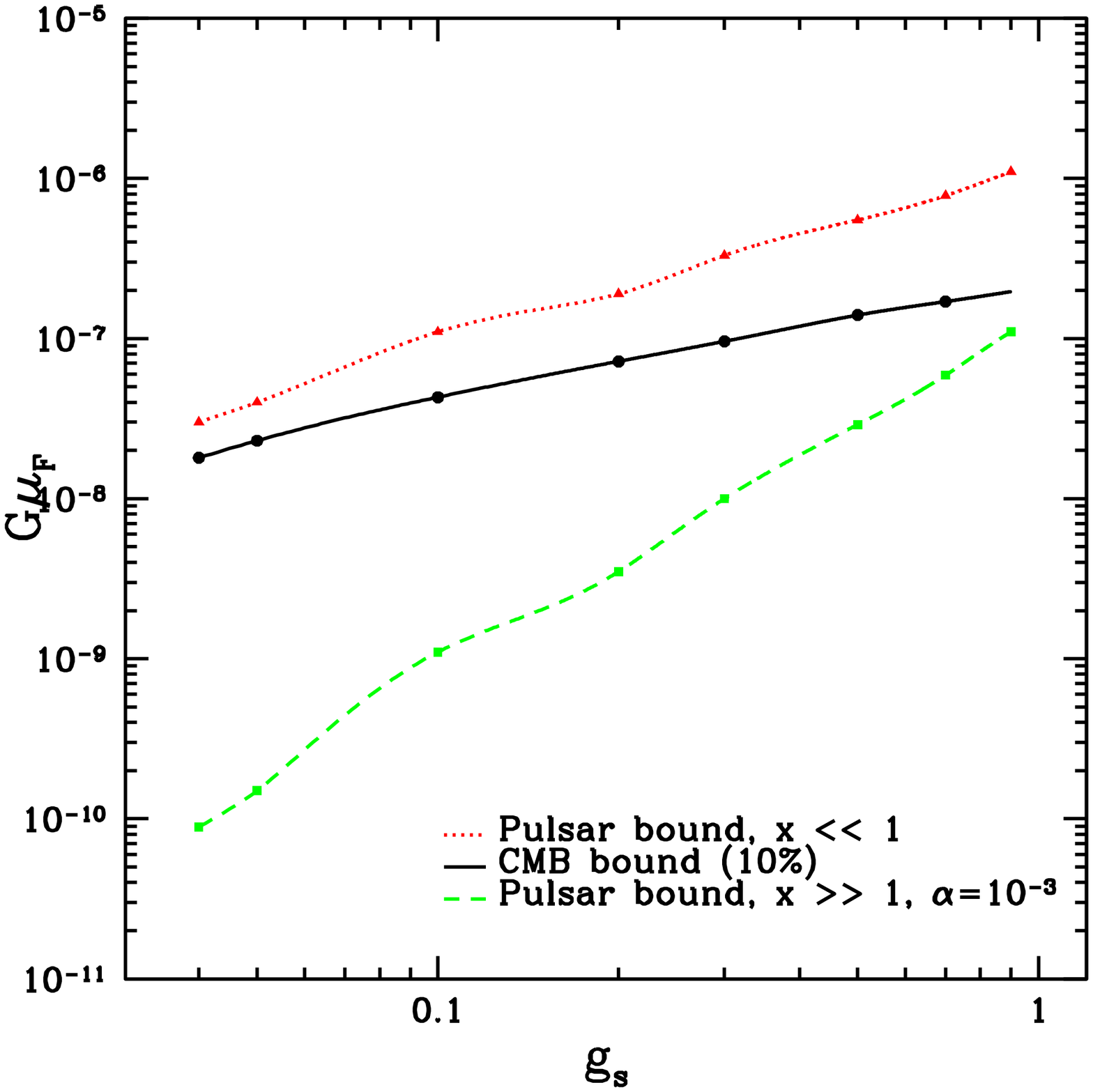}
\includegraphics[scale=0.4]{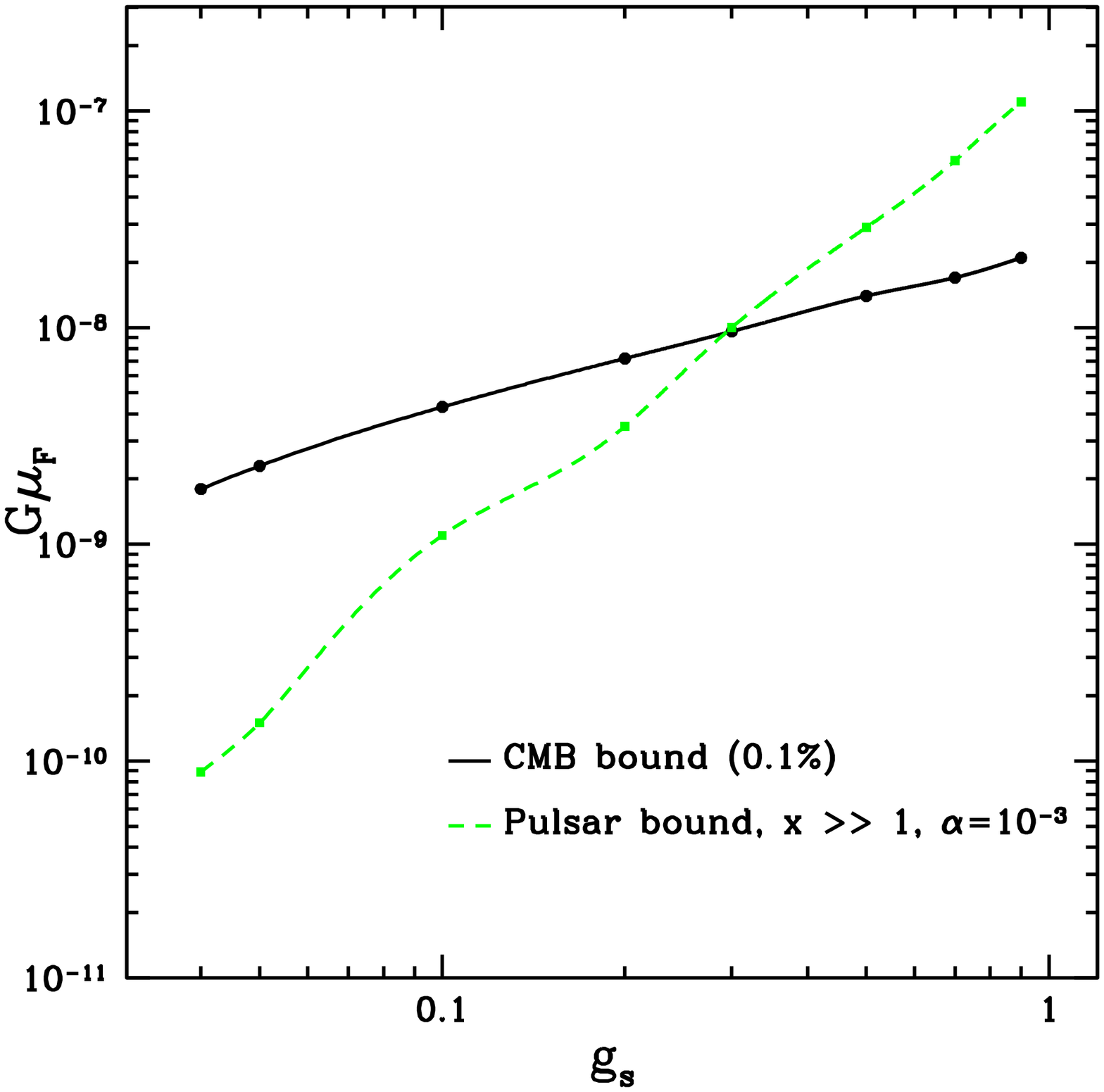}
\caption{{\it Left:} Current bounds on $\mu_F$ and $g_s$ from CMB (solid-black line) and pulsars (short-dash-green and dot-red line). {\it Right:} The forecasted bound on $\mu_F$ and $g_s$ from CMB based on future BB corresponding to $0.1\%$ in strings (solid-black line) together with the bounds from pulsars for the case $x_i \gg 1$ with $\alpha=0.001$ (short-dash-green line).}
\label{fig:bounds}
\end{figure}

As mentioned in the previous section, the amplitudes of CMB two-point correlations do not separately constrain the string tensions and their densities. Instead, they constrain a combination of $\mu_i$ and $\xi_i$ given by Eq.~(\ref{clstringN}). In particular, they do not differentiate between dense networks of light strings and rare heavy strings. In the case of FD networks, the relative abundances of different types of strings are controlled by the string coupling $g_s$, with the bound on the CMB normalisation leading to different values of $\mu_F$ for different values of $g_s$. Thus, the requirement that strings contribute no more than $10\%$ of the total CMB TT power can be translated into a joint constraint on $\mu_F$ and $g_s$ shown with a solid line in Fig.~\ref{fig:bounds}. As mentioned in the previous section, the dependence of this bound on the detailed position of the peak in the spectrum is relatively weak. The derived bounds on $\mu_F$ and $g_s$ depend more directly on the normalisation of the spectra given by Eq.~(\ref{clstringN}).

The degeneracy between $\mu_i$ and $\xi_i$, or in the case of FD networks, between $\mu_F$ and $g_s$, can be partially broken if other types of observations become available. One example would be a measurement of the position of the string induced bump in the BB spectrum. Based on the results in Sec.~\ref{sec:cmb}, if a peak is found at $\ell_{\rm peak} \approx 610 \pm 50$, that would strongly disfavour $g_s>0.1$, while $\ell_{\rm peak} \approx 750 \pm 50$ would disfavour $g_s< 0.3$. In the future, more sophisticated simulations of CMB from MTSN based on specific compactifications can, in principle, make more accurate predictions for the dependence of the BB peak position on $g_s$.

A second way to break the degeneracy between $\mu_F$ and $g_s$ is to combine bounds from the CMB with bounds on gravity waves (GW) emitted by strings, such as those coming from pulsar timing experiments \cite{Jenet} (for other recent constraints coming from pulsar timing and lensing see \cite{Pshirkov}) and direct GW searches by LIGO \cite{ligo}. The GW bounds constrain the energy density in strings approximately given by the combination $\mu/\xi^2$ for each type of string. The fact that the functional dependences of the GW and CMB bounds on $\mu_i$ and $\xi_i$ are different implies that by combining the two probes one can, in principle, reduce the degeneracy between $\mu_F$ and $g_s$.

To illustrate this point, we follow the procedure presented in \cite{Battye:2010xz}, where the authors calculated the bounds on the cosmic string tension from pulsar timing \cite{Jenet} (note that they are stronger than the ones coming from LIGO). For a network of a single type of string, the formula is \cite{CBS}
\begin{equation}
\label{pulsareq}
\Omega_{\rm g}h^2=1.17\times 10^{-4}{G\mu\over c^2}\left({1-{\langle v_{\rm rad}^2\rangle/ c^2}\over \xi_{\rm rad}^2\Omega_{\rm m}}\right){(1+1.4x)^{3/2}-1\over x}\,,
\end{equation}
where $x=\alpha c^2/(\Gamma \, G\mu)$, $\alpha$ is the loop production size, and $\Omega_{\rm m}$ is the total matter density relative to the critical density. They use parameters measured from the Nambu simulations to give $\xi_{\rm rad}$, $\langle v_{\rm rad}^2\rangle$ and set $\Omega_{\rm m}=0.3$, $\Gamma = 60$. We generalise this formula to include the three types of string that dominate the energy density of the FD network, namely the F, D and FD strings. Setting $c=1$ we write
\begin{equation}
\label{pulsareq_gen}
\Omega_{\rm g}h^2=1.17\times 10^{-4}\displaystyle\sum_{i=1}^{3}{G\mu_i}
\left({1-{\langle v_{\rm rad,i}^2\rangle}
\over \xi_{\rm rad,i}^2\Omega_{\rm m}}\right){(1+1.4x_i)^{3/2}-1\over x_i}\,,
\end{equation}
where $i=1,2,3$ correspond to the F, D and FD string respectively. Also, $x_i=\alpha/(\Gamma \, G\mu_i)$, and for simplicity we take the $\alpha$ and $\Gamma$ parameters to be the same for all types of string. For a given value of $g_s$, we use the extended VOS model of Sec.~\ref{sec:scaling} to determine the values of $\xi_i$. With those in hand, given the bound  $\Omega_{\rm g}h^2 < 2 \times 10^{-8}$, which is the most reliable published limit \cite{Jenet}, we can use the relations $\mu_D=\mu_F/g_s$ and $\mu_{FD}=\mu_F\sqrt{g_s^{-2}+1}$ to find the bound on $G\mu_F$.
As in \cite{Battye:2010xz} we consider two limiting cases of $x_i \ll 1$ and $x_i \gg 1$ \\\\
\textbf{Case a: $x_i \ll 1$:} In this case, eq.~(\ref{pulsareq_gen}) becomes
\begin{equation}
\Omega_{\rm g}h^2=\frac{3}{2}\cdot1.4\cdot1.17\times 10^{-4}\displaystyle\sum_{i=1}^{3}{G\mu_i}
\left({1-{\langle v_{\rm rad,i}^2\rangle}
\over \xi_{\rm rad,i}^2\Omega_{\rm m}}\right),
\end{equation}
so it is independent of $\alpha$ (as in \cite{Battye:2010xz}). 
The corresponding joint constraint on $\mu_F$ and $g_s$ is shown with red dotted line in Fig.~\ref{fig:bounds} \\\\
\textbf{Case b:  $x_i \gg 1$:} In this case, eq.~(\ref{pulsareq_gen}) becomes
\begin{equation}
\Omega_{\rm g}h^2=1.4^{3/2}\cdot1.17\times 10^{-4}
\displaystyle\sum_{i=1}^{3}{G\mu_i}\left({1-{\langle v_{\rm rad,i}^2\rangle}
\over \xi_{\rm rad,i}^2\Omega_{\rm m}}\right)x_i^{1/2}.
\end{equation}
As in \cite{Battye:2010xz}, substituting $x_i=\alpha/(\Gamma \, G\mu_i)$ results in a $1/\alpha$ dependence of $\mu_F$.
The corresponding joint constraint on $\mu_F$ and $g_s$ for $\alpha=0.001$ is shown with a short-dash green line in Fig.~\ref{fig:bounds}. Note that the chosen value of $\alpha$ gives a bound of $G\mu < 5 \cdot 10^{-8}$ for a `usual' cosmic string with $\xi_{rad}=0.13$ and $v=0.65$ \cite{Battye:2010xz}.

In the limit $x_i \ll 1$, the bounds on $\mu_F$ are weaker than the ones coming from the CMB. On the other hand,
for $x_i \gg 1$ the bounds can be much stronger, because of the $1/\alpha$ dependence. However, the uncertainties
in the upper bound on $\Omega_{\rm g}h^2$ in this limit are much more severe \cite{Battye:2010xz}.
Note that the shapes of the two pulsar bound curves in Fig.~\ref{fig:bounds} for the two limits ($x_i \ll 1$ and $x_i \gg 1$) are different because the dependence on $\mu_i$ is different.

Future probes of the B-mode signal will be able to constrain cosmic strings down to a $0.1\%$ contribution to TT, which translates into an order of magnitude tighter bound on $\mu_F$. In the right panel of Fig.~\ref{fig:bounds} we show the corresponding expected bounds from the CMB.

In the context of specific brane inflation scenarios, the cosmic string based constraints on $g_s$ and $\mu_F$ would need to be considered in jointly with other predictions of brane inflation, such as the spectral index $n_s$, the tensor-to scalar ratio $r$ and the tensor index $n_T$, as well as possible departures from Gaussianity in the distribution of the primordial fluctuations. We expect that such a comprehensive approach can lead to non-trivial constraints on details of brane inflation models, and more generally, the fundamental parameters of string theory.

\section{conclusions}
\label{sec:conc}

Using the MTSN scaling model developed in \cite{Avgoustidis:2009ke}, we have studied the evolution of a network of cosmic superstrings with the aim of identifying characteristic trends in their scaling properties at different values of the fundamental string coupling $g_s$. One particularly interesting trend that we have discovered is in the so-called power spectrum density, which controls the amplitude of the two-point function of the string stress-energy. As $g_s \rightarrow 1$ we see it is  dominated by populous light F and D strings where as as $g_s$ decreases to smaller values  rare heavy D strings come to dominate.

Using the MTSN scaling model in CMBACT, we evaluated the contribution of the FD networks to the CMB temperature and polarisation spectra. We found that the difference between the scaling patterns at high and low values of $g_s$ are manifest as a shift in position of the peak in the B-mode spectrum. In the one-scale approximation adopted in this work, the correlation length is equal to the average inter-string distance, thus string networks of higher (lower) number density have smaller (larger) correlation lengths. The correlation lengths, along with the string velocities, determine the position of the peak. This points to the possibility of constraining $g_s$ with future CMB experiments, possibly providing us with the first opportunity to date to constrain this fundamental parameter of string theory with observations.

Most observables, at least those that rely on averaged properties of string networks, constrain a combination of the string tension $\mu$ and their number density, determined by $\xi^{-2}$. Thus, in principle, it is quite hard to distinguish between the effect of many light strings vs a few heavy ones. Measuring a particular peak position in the B-mode spectrum would be one way to partially break this degeneracy, as we have discussed in this work. In addition, one can explore the fact that different observables constrain different combinations of $\mu$ and $\xi$. For instance, while the amplitude of the CMB spectra is determined by $(\mu / \xi)^2$, the energy density of strings is proportional to $\mu / \xi^2$. In this paper, we have shown how this difference can be explored in the case of FD networks to partially break the degeneracy between the fundamental string tension $\mu_F$ and the coupling $g_s$ by combining the CMB constraints with those from bounds on gravity waves (GW).

The main trends we have identified in this paper are largely independent of many of the details of the underlying string theory model, as well as the assumptions that went into the CMB calculation and the predictions for GW. However, follow up studies are needed in several directions in order to make firmer quantitative predictions. For instance, we need to improve our understanding of the string interaction rates for different choices of $g_s$ and $w$, especially in the non-perturbative regime. At very small string couplings, the density of the dominant species becomes so high that the one-scale approximation is almost guaranteed to break down. In such cases, a more sophisticated model is needed to properly describe the scaling of the network and its prediction for the CMB spectra. Whether it will be possible to measure a peak at high $\ell$ in the B-mode spectrum is another interesting question which is addressed in an upcoming publication \cite{MossPogosian}. It will depend strongly on the resolution and sensitivity of the experiments, as well as our ability to clean the contribution from weak lensing. The GW bounds on FD strings depend on the presence of cusps \cite{Davis:2008kg}, and on the loop size distribution, which is not fully understood at present, and should be revisited in the light of additional GW signatures coming from $Y$-junctions \cite{Binetruy:2010cc}. Also, while in all string models considered so far the B-mode from the ordinary strings is sourced predominately by vector modes, the tensor modes (i.e. large scale GW) were never properly worked out (since it requires accounting for the backreaction). It is our hope that the potentially very exciting opportunity for testing fundamental theory based on the general trends identified in this work will serve as an additional motivation for pursuing the remaining open questions.

\section{Acknowledgments} 
We acknowledge financial support from the Royal Society (EJC), the University of Manchester (AP), the Centre for Theoretical Cosmology in Cambridge (AA), NSERC Discovery Grant (LP), and CNRS (DS and LP). LP is grateful to the University of Nottingham and APC (Paris) for hospitality. DS is grateful to Simon Fraser University for hospitality whilst this work was in progress. AP is grateful to Simon Fraser University for hospitality during the early stages of this work and to the University of Nottingham for financial support while most of this work was being completed. AA is grateful to the Institute of Cosmos Sciences, University of Barcelona, for hospitality and support. The authors would like to thank Aaron Berndsen for help with numerics, Richard Battye, Paul Shellard and Donough Regan for useful discussions, and Mark Jackson for numerous clarifying discussions regarding the string theory aspect of this work in section \ref{superparams}.

\section*{Appendix: Approximate analytical solutions}\label{analytics}

A surprisingly significant amount of analytic progress can be made with just a few assumptions about the general forms of the scaling solutions. Let us first revisit the usual one-scale case. In Sec.~\ref{sec:vos} we have seen that in the VOS model the scaling constant values of the rms velocity $v_{\rm rms}$ and $\xi$ are given by Eqs.~(\ref{vosscaling1}) and (\ref{vosscaling2}) when $k$ is given by Eq.~(\ref{k-eqn}).
To obtain an estimate of their values, let us make an assumption that  $v^2 \simeq 1/2$, which is based on evidence from the numerical solutions. It suggests that a perturbative solution to the equations could be obtained by writing
\be \label{epsilon-def}
v^2 = \frac{1}{2} - \epsilon
\ee
where $\epsilon \ll 1$.  It then follows from Eq.~(\ref{k-eqn}) that $k \simeq  6\sqrt{2}\epsilon/\pi$ and hence from Eq.~(\ref{vosscaling1}) 
\be
\xi \simeq \frac{6\epsilon}{\pi \beta}
\label{xieps}
\ee
indicating that $\xi = {\cal O}(\epsilon)$. In fact, from Eq.~(\ref{vosscaling1}) and Eq.~(\ref{k-eqn}) we have 
\be
v^2={\tilde{\beta} (1- 8v^6) \over 1- 8v^6 + \tilde{C} (1+8v^6)}
\label{v-eqn}
\ee
where
\be
\tilde{C} = \frac{\pi \tilde{c}}{2\sqrt{2}} , \qquad \tilde{\beta}=\frac{1-\beta}{\beta}
\ee

This can be solved for $\epsilon$ using Eq.~(\ref{epsilon-def}), leading to solutions in the radiation ($\beta = 1/2$) and matter  ($\beta = 2/3$) eras 
\be
\epsilon_{\rm rad} = {\tilde{C} \over 3} ~~{\rm and}~~ \epsilon_{\rm mat} = \sqrt{{\tilde{C}\over 6}}.
\label{ep-rad}
\ee
Note this implies that  $\tilde{C}  = {\cal O}(\epsilon)$ during the radiation era, where as in the matter era $\tilde{C}  = {\cal O}(\epsilon^2)$. To give a flavour of the values these take, if $\tilde{c}_r \simeq 0.23$  (i.e. $\tilde{C}_r  = 0.26$) in the radiation era and $\tilde{c}_m \simeq 0.18$ (i.e. $\tilde{C}_m  = 0.2$) in the matter era, then
\be
\epsilon_{\rm rad} = {\tilde{C}_r \over 3} = 0.09 \to \xi_{\rm rad} \simeq 0.34 ~ {\rm and}~ v_{\rm rad}^2 \simeq 0.41 \qquad (v_{\rm rad} \sim 0.64)
\label{ep-rad1}
\ee
whilst
\be
\epsilon_{\rm mat} = \sqrt{{\tilde{C}_m \over 6}} = 0.18 \to \xi_{\rm mat} \simeq 0.5 ~ {\rm and}~ v_{\rm mat}^2 \simeq 0.32 \qquad (v_{\rm mat} \sim 0.57)
\label{ep-mat1}
\ee 

Let us now consider the MTSN model of Sec.~\ref{multimu} with parameters of Sec.~\ref{superparams}. The tension of a $(p,q)$ string is given by Eq.~(\ref{pqtension}) and,
for simplicity, we consider the first three tensions for the F, D and the $(1,1)$ string:
\bq
\mu_1 \equiv \mu_{(1,0)} &=& \mu_F
\\
\mu_2 \equiv \mu_{(0,1)} &=& \mu_F g_s^{-1}
\\
\mu_3 \equiv \mu_{(1,1)} &=& \mu_F \sqrt{1 + g_s^{-2}} \ .
\eq
Of course many other types of string could be considered, however in practice only these three string types will be used as they will be the most populous.  For $g_s=1$, $\mu_1=\mu_2 = \mu_3/\sqrt{2}$,  
where as if $g_s \ll 1$ then $\mu_1 \ll \mu_2 \sim \mu_3$. 

Next, we consider the evolution Eqs.~(\ref{rho_idtgen}-\ref{v_idtgen}) in Sec.~\ref{multimu} with the coefficients $b_{ab}^i$ set to zero. Then, as with the case of a single type of string, the velocity $v_i$ of the $i$th string satisfies 
\bq
    \dot v_i = (1-v_i^2)\left[\frac{k_i}{L_i}-2\frac{\dot a}{a}v_i 
    \right] \, . \label{v_idtgen1}     
\eq
with $k_i$ given by Eq.~(\ref{ki-eqn}).
Generalising the definition of $\epsilon$ in Eq.~(\ref{epsilon-def}) we write for the velocity of the $i$th string
\be
v_i^2 = \frac{1}{2} - \epsilon_i
\ee
where $\epsilon_i \ll 1$. Eq.~(\ref{ki-eqn}) then leads to $k_i \simeq  6\sqrt{2}\epsilon_i/\pi$, hence from Eq.~(\ref{v_idtgen1}) it follows that at scaling 
\be
\xi_i \simeq  \frac{6\epsilon_i}{\pi \beta} =  {\cal O}(\epsilon)
\label{xiieps}
\ee
providing the analogous solutions to Eq.~(\ref{xieps}). Using $\ell_{ij}^k$ given by Eq.~(\ref{zipper-length}),
the equation of motion Eq.~(\ref{rho_idtgen}) for the string density 
reduces to
\be
2 - 2\beta(1+v_i^2) - \frac{c_i 
    v_i}{\xi_i} = \sum_{a,k} \frac{d_{ia}^k   
    \bar v_{ia} \xi_i }{(\xi_i + \xi_a) \xi_a} - \sum_{b,\,a\le b}   
     \frac{d_{ab}^i   
    \bar v_{ab} \xi_i^2 }{(\xi_a + \xi_b) \xi_a \xi_b}.
    \label{reduces}
\ee
Of course, if the interaction term $d_{ab}^c=0$, then this corresponds to the case of the single type of string. Now, since 
\be
\bar{v}_{ab} = \sqrt{v_a^2 + v_b^2} = 1 + {\cal O}(\epsilon),
\ee
it follows that by considering the order of each term  in Eq.~(\ref{reduces}), we require 
\be
d_{ab}^i =  {\cal O}(\epsilon) ~~{\rm radiation~era}
\ee
and 
\be
d_{ab}^i =  {\cal O}(\epsilon^2) ~~{\rm matter~era}\,.
\ee
For three string types, Eq.~(\ref{reduces}) becomes
in terms of $\epsilon_i$,
\be
6\epsilon_1 \left(\frac{1}{\beta} - \frac{3}{2} \right)
+
6\epsilon_1^2 - \tilde{C}_1(1-\epsilon_1) = \frac{\pi \epsilon_1^2 }{2} 
\left( \frac{A}{\epsilon_2(\epsilon_1 + \epsilon_2)} +  \frac{C}{\epsilon_3(\epsilon_1 + \epsilon_3)} - \epsilon_1 \frac{B}{\epsilon_2 \epsilon_3 (\epsilon_2 + \epsilon_3)} \right)
\label{str1e}
\ee
\be
6\epsilon_2 \left(\frac{1}{\beta} - \frac{3}{2} \right)
+
6\epsilon_2^2 - \tilde{C}_2(1-\epsilon_2) =\frac{\pi \epsilon_2^2}{2} 
     \left(   \frac{B}{\epsilon_3(\epsilon_2 + \epsilon_3)} + \frac{A}{\epsilon_1(\epsilon_1 + \epsilon_2)} - \epsilon_2 \frac{C}{\epsilon_1 \epsilon_3 (\epsilon_1 + \epsilon_3)}\right)
\label{str2e}
\ee
\be
6\epsilon_3 \left(\frac{1}{\beta} - \frac{3}{2} \right)
+
6\epsilon_3^2 - \tilde{C}_3(1-\epsilon_3) =\frac{\pi \epsilon_3^2}{2} \left( \frac{C}{\epsilon_1(\epsilon_1 + \epsilon_3)} +  \frac{B}{\epsilon_2(\epsilon_2 + \epsilon_3)} - \epsilon_3 \frac{A}{\epsilon_1 \epsilon_2 (\epsilon_1 + \epsilon_2)}\right)
\label{str3e}
\ee
where 
\be
A = d_{12}^{3}\bar v_{12}, \qquad B = d_{23}^{1}\bar v_{23}, \qquad C = d_{13}^{2}\bar v_{13} \ .
\ee

To solve the system of Eqs. (\ref{str1e} - \ref{str3e}) we start by considering the results from the numerical simulations which were in the radiation dominated era ($\beta = 1/2$). For $g_s \ll 1$, and for almost all values of $w$, we find that 
\be
c_1 \ll c_2 \simeq c_3~~{\rm and}~~A \simeq C \ll B \ ,
\ee
as seen, for example, in Table~\ref{Tablecisdijks}. Then, as a first approximation we set $A=C=0$, $c_1=0$, $\bar v_{ab} =1$ and $\beta = \frac12$.

In that case Eqs. (\ref{str1e})-(\ref{str3e}) reduce to
\be
3 \epsilon_1 - \tilde{C}_1
 = - \frac{\pi}{2} 
\epsilon_1^2 
\left( \epsilon_1 \frac{B}{\epsilon_2 \epsilon_3 (\epsilon_2 + \epsilon_3)} \right)
\label{str1g}
\ee
\be
3\epsilon_2 - \tilde{C}_2=\frac{\pi \epsilon_2^2}{2} 
     \left(   \frac{B}{\epsilon_3(\epsilon_2 + \epsilon_3)} \right)
\label{str2g}
\ee
\be
3\epsilon_3 - \tilde{C}_3=\frac{\pi \epsilon_3^2}{2}  \left(  \frac{B}{\epsilon_2(\epsilon_2 + \epsilon_3)}\right)
\label{str3g}
\ee
As we have set $c_2 \simeq c_3$ then $\tilde{C}_2 = \tilde{C}_3$ implying that 
$\epsilon_2=\epsilon_3$ with Eq.~(\ref{str2g}) and Eq.~(\ref{str3g}) reducing to 
\be
\epsilon_2=\epsilon_3 = \left( \frac{\tilde{C}_2 + B \pi/4 }{3} \right) > \bar{\epsilon}_2
\label{eps2a}
\ee
where $\bar{\epsilon}_2$ corresponds to the solution Eq.~(\ref{ep-rad}) in which the strings are not interacting. The effect of $B$ (with $A=C=0$) is therefore to {\it reduce} the velocity of the D and $(1,1)$ string, but to {\it increase} their correlation length and thus {\it decrease} their number density. 
From Eq.~(\ref{str1g}) 
\be
\label{eps1a}
3 \epsilon_1 - \tilde{C}_1 = - \frac{\epsilon_1^3 B \pi}{4 \epsilon_2^3}
\ee
where $\epsilon_2$ is given in Eq.~(\ref{eps2a}). Since $c_1 \ll c_2$ we expect $\epsilon_1 \ll \epsilon_2$ hence to a first approximation the solution for the F string is that corresponding to a network of only one type of string, namely 
 \be
\label{eps1b}
\epsilon_1 = \frac{\tilde{C}_1}{3}. 
\ee
implying that the velocity of this string will be very close to $1/\sqrt{2}$ and $\xi_1$ will again be small. 
In terms of $\xi_i$ and the original $c_i$ we have 
\be
\label{xi-predict}
\xi_1 = \sqrt2 c_1\,,~~~\xi_2 = \xi_3 \simeq \sqrt2 c_2 + d_{23}^{1}
\ee
The comparison with the simulations is very encouraging. For example for the case $w=0.1$, the prediction for $\xi_1$ in Eq~(\ref{xi-predict}) matches the simulations exactly for $0.04 < g_s < 0.5$. The approximation we have adopted has forced $\xi_2=\xi_3$ and the predictions match very well the numerical solutions for $\xi_3$ for values of $g_s < 0.3$, differing by no more than 7\%. It fails though to allow for the observed deviation between $\xi_2$ and $\xi_3$ although this could well be accommodated for by an iterative approach to the solution.

\end{document}